\documentclass[preprint,showpacs,preprintnumbers,amsmath,amssymb]{revtex4}
\usepackage{graphicx}
\usepackage{dcolumn}
\usepackage{bm}

\newcommand\lab[1]{\label{eq:#1}}

\newcommand\br{\begin{eqnarray}}
\newcommand\er{\end{eqnarray}}
\newcommand\be{\begin{equation}}
\newcommand\ee{\end{equation}}

\newcommand\bc{\begin{center}}
\newcommand\ec{\end{center}}

\newcommand\partder[2]{\frac{{\partial {#1}}}{{\partial {#2}}}}

\newcommand\PRL[3]{\textsl{Phys. Rev. Lett.} \textbf{#1}, #3 (#2)}

\newcommand\PRD[3]{\textsl{Phys. Rev.} \textbf{D#1}, #3 (#2)}

\newcommand\PLB[3]{\textsl{Phys. Lett.} \textbf{#1B}, #3 (#2)}
\newcommand\CQG[3]{\textsl{Class. Quantum Grav.} \textbf{#1}, #3 (#2)}

\newcommand\AoP[3]{\textsl{Ann. of Phys.} \textbf{#1}, #3 (#2)}
\newcommand\RMP[3]{\textsl{Rev. Mod. Phys.} \textbf{#1}, #3 (#2)}

\newcommand\IJMPA[3]{\textsl{Int. J. Mod. Phys.} \textbf{A#1}, #3 (#2)}
\newcommand\IJMPD[3]{\textsl{Int. J. Mod. Phys.} \textbf{D#1}, #3 (#2)}

\newcommand\JPA[3]{\textsl{J. Physics} \textbf{A#1}, #3 (#2)}

\newcommand\MPLA[3]{\textsl{Mod. Phys. Lett.} \textbf{A#1}, #3 (#2)}

\begin{document}


\title{Connecting The Non-Singular Origin of the Universe, The Vacuum Structure and The Cosmological Constant Problem}

\author{Eduardo I. Guendelman}
\email{guendel@bgu.ac.il} \affiliation{ Physics Department, Ben
Gurion University of the Negev, Beer Sheva 84105, Israel}
\author{Pedro Labra\~{n}a}
\email{plabrana@ubiobio.cl, plabrana@icc.ub.edu} \affiliation{Departamento de F\'{i}sica, Universidad del B\'{i}o-B\'{i}o, Casilla 5-C, Concepci\'on, Chile and \\
Departament d'Estructura i Constituents de la Mat\`{e}ria, Institut
de Ci\`{e}ncies del Cosmos, Universitat de Barcelona, Diagonal 647,
08028 Barcelona, Spain.}

\begin{abstract}
We consider a non-singular origin for the Universe starting from an
Einstein static Universe, the so called ``emergent universe"
scenario, in the framework of a theory which uses two volume
elements $\sqrt{-{g}}d^{4}x$ and $\Phi d^{4}x$, where $\Phi $ is a
metric independent density, used as an additional measure of
integration. Also curvature, curvature square terms and for scale
invariance a dilaton field $\phi$  are considered in the action. The
first order formalism is applied.  The integration of the equations
of motion associated with the new measure gives rise to the
spontaneous symmetry breaking (S.S.B) of scale invariance (S.I.).
After S.S.B. of S.I., it is found that a non trivial potential for
the dilaton is generated. In the Einstein frame we also add a
cosmological term that parametrizes the zero point fluctuations. The
resulting effective potential for the dilaton contains two flat
regions, for $\phi \rightarrow \infty$ relevant for the non-singular
origin of the Universe, followed by an inflationary phase and $\phi
\rightarrow -\infty$, describing our present Universe. The dynamics
of the scalar field becomes non linear and these non linearities
produce a non trivial vacuum structure for the theory and are
responsible for the stability of some of the emergent universe
solutions, which exists for a parameter range of values of the
vacuum energy in $\phi \rightarrow -\infty$, which must be positive
but not very big, avoiding the extreme fine tuning required to keep
the vacuum energy density of the present universe small. The non
trivial vacuum structure is crucial to ensure the smooth transition
from the emerging phase, to an inflationary phase and finally to the
slowly accelerated universe now. Zero vacuum energy density for the
present universe defines the threshold for the creation of the
universe.
\end{abstract}

\pacs{98.80.Cq, 04.20.Cv, 95.36.+x}


\maketitle
\section{Introduction}
One  of the most important and intriging issues of modern physics is
the so called ``Cosmological Constant Problem" \cite{CCP1, CCP2,
CCP3, CCP4, CCP5}, (CCP), most easily seen by studying the
apparently uncontrolled behaviour of the zero point energies, which
would lead to a corresponding equally uncontrolled vacuum energy or
cosmological constant term. Even staying at the classical level, the
observed very small cosmological term in the present universe is
still very puzzling.

Furthermore, the Cosmological Constant Problem has evolved from the
``Old Cosmological Constant Problem",  where  physicist were
concerned with explaining why the observed vacuum energy density of
the universe is exactly zero, to different type of CCP since the
evidence for the accelerating universe became evident, for reviews
see \cite{AU1, AU2}.  We have therefore since the discovery of the
accelerated universe a  ``New Cosmological Constant Problem"
\cite{W2},  the problem is now not to explain zero, but to explain a
very small vacuum energy density.

This new situation posed by the discovery of a very small vacuum
energy density of the universe means that getting a zero vacuum
energy density for the present universe is definitely not the full
solution of the problem, although it may be a step towards its
solution.

One point of view to the CCP that has been popular has been to
provide a bound based on the ``anthropic principle"
\cite{Anthropic}. In this approach, a too large Cosmological
Constant will not provide the necessary conditions required for the
existence of life, the anthropic principle provides then an upper
bound on the cosmological constant.

One problem with this approach is for example that it relies on our knowledge of life as we know it and ignores the possibility that other life forms could be possible, for which other (unknown) bounds would be relevant, therefore the reasoning appears by its very nature subjective, since of course if the observed cosmological constant will be different, our universe will be different and this could include different kind of life that may be could have adjusted itself to a higher cosmological constant of the universe. But even accepting the validity of anthropic considerations, we still do not understand why the observed vacuum energy density must be positive instead of possibly a very small negative quantity. Accepting the anthropic explanation means may be also giving up on discovering important physics related to the  CCP and this may be the biggest objection.

Nevertheless, the idea of associating somehow restrictions on the
origin of the universe with the cosmological constant problem seems
interesting. We will take on this point of view, but leave out the
not understood concept of life out from our considerations. Instead,
we will require, in a very specific framework, the non-singular
origin of the universe. The advantage of this point of view is that
it is formulated in terms of ideas of physics alone, without
reference to biology, which unlike physics, has not reached the
level of an exact science. Another interesting consequence is that
we can learn that of a non-singularly created universe may not have
a too big cosmological constant, an effect that points to a certain
type of gravitational suppression of UV divergences in quantum field
theory.

In this respect, one should point out that even in the context of
the inflationary scenario \cite{Inflation1, Inflation2, Inflation3,
Inflation4} which solves many cosmological problems, one still
encounters the initial singularity problem which remains unsolved,
showing that the universe necessarily had a singular beginning for
generic inflationary cosmologies \cite{singularities1,
singularities2, singularities3, singularities4, singularities5}.

Here we will adopt the very attractive ``Emergent Universe"
scenario, where those conclusions concerning singularities can be
avoided \cite{emerging1, emerging2, emerging3, emerging4, emerging5,
emerging6, emerging7, emerging8, emerging9}. The way to escape the
singularity in these models is to violate the geometrical
assumptions of these theorems, which assume i) that the universe has
open space sections ii) the Hubble expansion is always greater than
zero in the past. In \cite{emerging1, emerging2} the open space
section condition is violated since closed Robertson Walker
universes with $k=1$ are considered and the Hubble expansion can
become zero, so that both i) and ii) are avoided.

In \cite{emerging1, emerging2} even models based on standard General
Relativity, ordinary matter and minimally coupled scalar fields were
considered and can provide indeed a non-singular (geodesically
complete) inflationary universe, with a past eternal Einstein static
Universe that eventually evolves into an inflationary Universe.

Those most simple models suffer however from instabilities,
associated with the instability of the Einstein static universe. The
instability is possible to cure by going away from GR, considering
non perturbative corrections to the Einstein's field equations in
the context of the loop quantum gravity \cite{emerging3}, a brane
world cosmology with a time like extra dimension \cite{emerging4,
emerging5} considering the Starobinski model for radiative
corrections (which cannot be derived from an effective action)
\cite{emerging6} or exotic matter \cite{emerging7}. In addition to
this, the consideration of a Jordan Brans Dicke model also can
provide a stable initial state for the emerging universe scenario
\cite{emerging8, emerging9}.

In this review we study a different theoretical framework where such
emerging universe scenario is realized in a natural way, where
instabilities are avoided and a succesfull  inflationary phase with
a graceful exit can be achieved. The  model we will use was studied
first in \cite{SIchile}, however, in the context of this model, a
few scenarios are possible. For example in the first paper on this
model  \cite{SIchile} a special choice of state to describe the
present state of our universe was made. Then in \cite{SICCP} a
different candidate for the vacuum that represents our present
universe was made. The way in which we best represents the present
state of the universe is crucial, since as it should be obvious, the
discussion of the CCP depends on what vacuum we take. In
\cite{SICCP} we expressed the stability and existence conditions for
the non-singular universe in terms of the energy of the vacuum of
our candidate for the present Universe. In \cite{SICCP}  a few typos
in \cite{SIchile} were corrected and also the discussion of some
notions discussed was improved  in \cite{SICCP} and more deeper
studies will be done  in  this review.

Indeed in  this review, all those topics will be further clarified,
in particular the vacuum structure of this model will be extended. A
very important new feature that will be presented in this review is
the existence of a ``kinetic vacuum", that produces a vacuum energy
state which is  degenerate with the vacuum choice made in
\cite{SICCP}, this degeneracy is analyzed and the dynamical role of
this kinetic vacuum in the evolution of the universe and the CCP is
analyzed.

We work in the context of a theory built along the lines of the two
measures theory (TMT). Basic idea is developed in \cite{TMT1a},
\cite{TMT1b}-\cite{TMT1r} \cite{TMT2}, \cite{TMT3a}-\cite{TMT3d},
\cite{TMT4a}-\cite{TMT4e}, \cite{TMT5} and more specifically in the
context of the scale invariant realization of such theories
\cite{TMT2}, \cite{TMT3a}-\cite{TMT3e}, \cite{TMT4a}-\cite{TMT4e},
\cite{TMT5}. These theories can provide a new approach to  the
cosmological constant problem and can be generalized to obtain also
a theory with a dynamical spacetime \cite{dyn}, furthermore, string
and brane theories, as well as brane world scenarios can be
constructed using Two Measure Theories ideas \cite{G16a}-\cite{G20}.
We should also point out that the Hodge Dual construction of
\cite{TMT1d} for supergravity constitutes in fact an example of a
TMT. The construction by Comelli \cite{TMT1e} where no square root
of the determinant of the metric is used and instead a total
divergence appears is also a very much related approach.

The two measure theories have many points of similarity with
``Lagrange Multiplier Gravity (LMG)''
\cite{Lim2010,Capozziello2010}. In LMG there is a Lagrange
multiplier field which enforces the condition that a certain
function is zero. For a comparison of one of these lagrange
multiplier gravity models with observations see
\cite{LabranaMultiplier}. In the two measure theory this is
equivalent to the constraint which requires some lagrangian to be
constant. The two measure model presented here, as opposed to the
LMG models of \cite{Lim2010,Capozziello2010} provide us with an
arbitrary constant of integration. The introduction of constraints
can cause Dirac fields to contribute to dark energy \cite{FDM} or
scalar fields to behave like dust like in \cite{Lim2010} and this
dust behaviour can be caused by the stabilization of a tachyonic
field due to the constraint, accompanied by a floating dark energy
component \cite{GSY, AG}. TMT models naturally avoid the 5th force
problem \cite{5th}.

We will consider a slight generalization of the TMT case, where, we
consider also the possible effects of zero point energy densities,
thus ``softly breaking" the basic structure of TMT for this purpose.
We will show how the stated goals of a stable emergent universe can
be achieved in the framework of the model and also how the stability
of the emerging universe imposes interesting constraints on the
energy density of the ground state of the theory as defined in this
paper: it must be positive but not very large, thus the vacuum
energy and therefore the term that softly breaks the TMT structure
appears to be naturally controlled. An important ingredient of the
model considered here is its softly broken conformal invariance,
meaning that we allow conformal breaking terms only though
potentials of the dilaton, which nevertheless preserve global scale
invariance. In another models for emergent universe we have studied
\cite{SICh2}, that rule of softly broken conformal invariance was
taken into account. It is also a perfectly consistent, but different
approach.

The review will be organized as follows: First we review the
principles of the TMT and in particular the model studied in
\cite{TMT2}, which has global scale invariance and how this can be
the basis for the emerging universe. Such model gives rise, in the
effective Einstein frame, to an effective potential for a dilaton
field (needed to implement an interesting model with global scale
invariance) which has a flat region. Following this, we look at the
generalization of this model \cite{TMT5} by adding a curvature
square or simply ``$R^{2}$ term" and show that the resulting model
contains now two flat regions. The existence of two flat regions for
the potential is shown to be consequence of the s.s.b. of the scale
symmetry.  We then consider the incorporation in the model of the
zero point fluctuations, parametrized by a cosmological constant in
the Einstein frame.  In this resulting model, there are two possible
types of emerging universe solutions, for one of those, the initial
Einstein Universe can be stabilized due to the non linearities of
the model, provided the vacuum energy density of the ground state is
positive but not very large. This is a very satisfactory results,
since it means that the stability of the emerging universe prevents
the vacuum energy in the present universe from being very large!.
The transition from the emergent universe to the ground state goes
through an intermediate inflationary phase, therefore reproducing
the basic standard cosmological model as well. We end with a
discussion section and present the point of view that the creation
of the universe can be considered as a ``threshold event" for zero
present vacuum energy density, which naturally gives a positive but
small vacuum energy density.

\section{Introducing a new measure}

The general structure of general coordinate invariant theories is taken usually as
\begin{equation}\label{1}
S_{1} = \int{L_{1}}\sqrt{-g} d^{4}x
\end{equation}
where $g =  det g_{\mu\nu}$.  The introduction of $\sqrt{-g}$ is
required since $d^{4}x$ by itself is not a scalar but the product
$\sqrt{-g} d^{4} x$ is a scalar. Inserting $\sqrt{-g}$, which has
the transformation properties of a density, produces a scalar action
$S_{1}$, as defined by Eq.(\ref{1}), provided $L_{1}$ is a scalar.

    In principle nothing prevents us from considering other densities instead of
$\sqrt{-g}$. One construction of such alternative ``measure of
integration", is obtained as follows:
 given 4-scalars $\varphi_{a}$ (a =
1,2,3,4), one can construct the density
\begin{equation}\label{2}
\Phi =  \varepsilon^{\mu\nu\alpha\beta}  \varepsilon_{abcd}
\partial_{\mu} \varphi_{a} \partial_{\nu} \varphi_{b} \partial_{\alpha}
\varphi_{c} \partial_{\beta} \varphi_{d}
\end{equation}
and consider in addition to the action $S_{1}$, as defined by
Eq.(\ref{1}), $S_{2}$, defined as
\begin{equation}\label{3}
S_{2} =  \int L_{2} \Phi d^{4} x
\end{equation}
$L_{2}$ is again some scalar, which may contain the curvature (i.e.
the gravitational contribution) and a matter contribution, as it can
be the case for $S_{1}$, as defined by Eq.(\ref{1}). For an approach
that uses four-vectors instead of four-scalars see
\citep{four-vector}.

    In the action $S_{2}$ defined by Eq.(\ref{3}) the measure carries degrees of freedom
independent of that of the metric and that of the matter fields. The most
natural and successful formulation of the theory is achieved when the
connection is also treated as an independent degree of freedom. This is
what is usually referred to as the first order formalism.

    One can consider both contributions, and allowing therefore both geometrical
objects to enter the theory and take as our action
\begin{equation}\label{e6}
S = \int L_{1} \sqrt{-g}d^{4}x + \int L_{2} \Phi  d^{4} x
\end{equation}

 Here $L_{1}$ and
$L_{2}$ are
$\varphi_{a}$  independent.

    We will study now the dynamics of a scalar field $\phi$ interacting
with gravity as given by the following action, where except for the potential terms
$U$ and $V$ we have conformal invariance, the potential terms
$U$ and $V$ break down this to global scale invariance.

\begin{equation}\label{e9}
S_{L} =    \int L_{1} \sqrt{-g}   d^{4} x +  \int L_{2} \Phi d^{4} x
\end{equation}
\begin{equation}\label{e10}
L_{1} = U(\phi)
\end{equation}

\begin{equation}\label{e11}
L_{2} = \frac{-1}{\kappa} R(\Gamma, g) + \frac{1}{2} g^{\mu\nu}
\partial_{\mu} \phi \partial_{\nu} \phi - V(\phi)
\end{equation}
\begin{equation}\label{e12}
R(\Gamma,g) =  g^{\mu\nu}  R_{\mu\nu} (\Gamma) , R_{\mu\nu}
(\Gamma) = R^{\lambda}_{\mu\nu\lambda}
\end{equation}
\begin{equation}\label{e13}
R^{\lambda}_{\mu\nu\sigma} (\Gamma) = \Gamma^{\lambda}_
{\mu\nu,\sigma} - \Gamma^{\lambda}_{\mu\sigma,\nu} +
\Gamma^{\lambda}_{\alpha\sigma}  \Gamma^{\alpha}_{\mu\nu} -
\Gamma^{\lambda}_{\alpha\nu} \Gamma^{\alpha}_{\mu\sigma}.
\end{equation}

The suffix $L$ in $S_{L}$ is to emphasize that here the curvature
appears only linearly. Here,  except for the potential terms $U$ and
$V$ we have conformal invariance, the potential terms $U$ and $V$
break down this to global scale invariance. Since the breaking of
local conformal invariance is only through potential terms, we call
this a ``soft breaking".

    In the variational principle $\Gamma^{\lambda}_{\mu\nu},
g_{\mu\nu}$, the measure fields scalars $\varphi_{a}$ and the
``matter" - scalar field $\phi$ are all to be treated as independent
variables although the variational principle may result in equations
that allow us to solve some of these variables in terms of others.

For the case the potential terms
$U=V=0$ we have local conformal invariance

\begin{equation}\label{e14}
g_{\mu\nu}  \rightarrow   \Omega(x)  g_{\mu\nu}
\end{equation}

and $\varphi_{a}$ is transformed according to
\begin{equation}\label{e15}
\varphi_{a}   \rightarrow   \varphi^{\prime}_{a} = \varphi^{\prime}_{a}(\varphi_{b})
\end{equation}

\begin{equation}\label{e16}
\Phi \rightarrow \Phi^{\prime} = J(x) \Phi
\end{equation}
 where $J(x)$  is the Jacobian of the transformation of the $\varphi_{a}$ fields.

This will be a symmetry in the case $U=V=0$ if
\begin{equation}\label{e17}
\Omega = J
\end{equation}
Notice that $J$ can be a local function of space time, this can be arranged by performing for the
$\varphi_{a}$ fields one of the (infinite) possible diffeomorphims in the internal $\varphi_{a}$ space.

 We can still retain a  global
scale invariance in model for very special exponential form for the $U$ and $V$ potentials. Indeed, if we perform the global
scale transformation ($\theta$ =
constant)
\begin{equation}\label{e18}
g_{\mu\nu}  \rightarrow   e^{\theta}  g_{\mu\nu}
\end{equation}
then (9) is invariant provided  $V(\phi)$ and $U(\phi)$ are of the
form  \cite{TMT2}
\begin{equation}\label{e19}
V(\phi) = f_{1}  e^{\alpha\phi},  U(\phi) =  f_{2}
e^{2\alpha\phi}
\end{equation}
and $\varphi_{a}$ is transformed according to
\begin{equation}\label{e20}
\varphi_{a}   \rightarrow   \lambda_{ab} \varphi_{b}
\end{equation}
which means
\begin{equation}\label{e21}
\Phi \rightarrow det( \lambda_{ab}) \Phi \\ \equiv \lambda
\Phi     \end{equation}
such that
\begin{equation}\label{e22}
\lambda = e^{\theta}
\end{equation}
and
\begin{equation}\label{e23}
\phi \rightarrow \phi - \frac{\theta}{\alpha}.
\end{equation}

We will now work out the equations of motion after introducing $V(\phi)$ and $U(\phi)$
and see how the integration of the equations of motion allows the spontaneous breaking of the
scale invariance.

    Let us begin by considering the equations which are obtained from
the variation of the fields that appear in the measure, i.e. the
$\varphi_{a}$
fields. We obtain then
\begin{equation}\label{e24}
A^{\mu}_{a} \partial_{\mu} L_{2} = 0
\end{equation}
where  $A^{\mu}_{a} = \varepsilon^{\mu\nu\alpha\beta}
\varepsilon_{abcd} \partial_{\nu} \varphi_{b} \partial_{\alpha}
\varphi_{c} \partial_{\beta} \varphi_{d}$. Since it is easy to
check that  $A^{\mu}_{a} \partial_{\mu} \varphi_{a^{\prime}} =
\frac{\delta aa^{\prime}}{4} \Phi$, it follows that
det $(A^{\mu}_{a}) =\frac{4^{-4}}{4!} \Phi^{3} \neq 0$ if $\Phi\neq 0$.
Therefore if $\Phi\neq 0$ we obtain that $\partial_{\mu} L_{2} = 0$,
 or that
\begin{equation}\label{e25}
L_{2} = \frac{-1}{\kappa} R(\Gamma,g) + \frac{1}{2} g^{\mu\nu}
\partial_{\mu} \phi \partial_{\nu} \phi - V = M
\end{equation}
where M is constant. Notice that this equation breaks spontaneously the global scale invariance of the theory,
since the left hand side has a non trivial transformation under the scale transformations, while the right
hand side is equal to $M$, a constant that after we integrate the equations is fixed, cannot be changed and therefore
for any $M\neq 0$ we have obtained indeed, spontaneous breaking of scale invariance.

We will see what is the connection now. As we will see, the connection appears in the original frame as a
non Riemannian object. However, we will see that by a simple conformal tranformation of the metric we can recover
the Riemannian structure. The interpretation of the equations in the frame gives then an interesting physical picture,
as we will see.

    Let us begin by studying the equations obtained from the variation of the
connections $\Gamma^{\lambda}_{\mu\nu}$.  We obtain then
\begin{equation}\label{e26}
-\Gamma^{\lambda}_{\mu\nu} -\Gamma^{\alpha}_{\beta\mu}
g^{\beta\lambda} g_{\alpha\nu}  + \delta^{\lambda}_{\nu}
\Gamma^{\alpha}_{\mu\alpha} + \delta^{\lambda}_{\mu}
g^{\alpha\beta} \Gamma^{\gamma}_{\alpha\beta}
g_{\gamma\nu}\\ - g_{\alpha\nu} \partial_{\mu} g^{\alpha\lambda}
+ \delta^{\lambda}_{\mu} g_{\alpha\nu} \partial_{\beta}
g^{\alpha\beta}
 - \delta^{\lambda}_{\nu} \frac{\Phi,_\mu}{\Phi}
+ \delta^{\lambda}_{\mu} \frac{\Phi,_           \nu}{\Phi} =  0
\end{equation}
If we define $\Sigma^{\lambda}_{\mu\nu}$    as
$\Sigma^{\lambda}_{\mu\nu} =
\Gamma^{\lambda}_{\mu\nu} -\{^{\lambda}_{\mu\nu}\}$
where $\{^{\lambda}_{\mu\nu}\}$   is the Christoffel symbol, we
obtain for $\Sigma^{\lambda}_{\mu\nu}$ the equation
\begin{equation}\label{e27}
    -  \sigma, _{\lambda} g_{\mu\nu} + \sigma, _{\mu}
g_{\nu\lambda} - g_{\nu\alpha} \Sigma^{\alpha}_{\lambda\mu}
-g_{\mu\alpha} \Sigma^{\alpha}_{\nu \lambda}
+ g_{\mu\nu} \Sigma^{\alpha}_{\lambda\alpha} +
g_{\nu\lambda} g_{\alpha\mu} g^{\beta\gamma} \Sigma^{\alpha}_{\beta\gamma}
= 0
\end{equation}
where  $\sigma = ln \chi, \chi = \frac{\Phi}{\sqrt{-g}}$.

    The general solution of Eq.(\ref{e28}) is
\begin{equation}\label{e28}
\Sigma^{\alpha}_{\mu\nu} = \delta^{\alpha}_{\mu}
\lambda,_{\nu} + \frac{1}{2} (\sigma,_{\mu} \delta^{\alpha}_{\nu} -
\sigma,_{\beta} g_{\mu\nu} g^{\alpha\beta})
\end{equation}\label{e30}
where $\lambda$ is an arbitrary function due to the $\lambda$ - symmetry
of the
curvature \cite{Lambda} $R^{\lambda}_{\mu\nu\alpha} (\Gamma)$,
\begin{equation}\label{e29}
\Gamma^{\alpha}_{\mu\nu} \rightarrow \Gamma^{\prime \alpha}_{\mu\nu}
 = \Gamma^{\alpha}_{\mu\nu} + \delta^{\alpha}_{\mu}
Z,_{\nu}
\end{equation}
Z  being any scalar (which means $\lambda \rightarrow \lambda + Z$).

    If we choose the gauge $\lambda = \frac{\sigma}{2}$, we obtain
\begin{equation}\label{e30}
\Sigma^{\alpha}_{\mu\nu} (\sigma) = \frac{1}{2} (\delta^{\alpha}_{\mu}
\sigma,_{\nu} +
 \delta^{\alpha}_{\nu} \sigma,_{\mu} - \sigma,_{\beta}
g_{\mu\nu} g^{\alpha\beta}).
\end{equation}

    Considering now the variation with respect to $g^{\mu\nu}$, we
obtain
\begin{equation}\label{e31}
\Phi (\frac{-1}{\kappa} R_{\mu\nu} (\Gamma) + \frac{1}{2} \phi,_{\mu}
\phi,_{\nu}) - \frac{1}{2} \sqrt{-g} U(\phi) g_{\mu\nu} = 0
\end{equation}
solving for $R = g^{\mu\nu} R_{\mu\nu} (\Gamma)$  from
Eq.(\ref{e31}) and introducing in Eq.\ref{e25}, we obtain
\begin{equation}\label{e32}
M + V(\phi) - \frac{2U(\phi)}{\chi} = 0
\end{equation}
a constraint that allows us to solve for $\chi$,
\begin{equation}\label{e33}
\chi = \frac{2U(\phi)}{M+V(\phi)}.
\end{equation}

    To get the physical content of the theory, it is best
consider variables that have well defined dynamical interpretation. The original
metric does not has a non zero canonical  momenta. The fundamental
variable of the theory in the first order formalism is the connection and its
canonical momenta is a function of $\overline{g}_{\mu\nu}$, given by,

\begin{equation}\label{e34}
\overline{g}_{\mu\nu} = \chi g_{\mu\nu}
\end{equation}

and $\chi$  given by Eq.(\ref{e33}). Interestingly enough, working
with $\overline{g}_{\mu\nu}$ is the same as going to the ``Einstein
Conformal Frame". In terms of $\overline{g}_{\mu\nu}$   the non
Riemannian contribution $\Sigma^{\alpha}_{\mu\nu}$ dissappears from
the equations. This is because the connection can be written as the
Christoffel symbol of the metric $\overline{g}_{\mu\nu}$ . In terms
of $\overline{g}_{\mu\nu}$ the equations of motion for the metric
can be written then in the Einstein form (we define
$\overline{R}_{\mu\nu} (\overline{g}_{\alpha\beta}) =$
 usual Ricci tensor in terms of the bar metric $= R_{\mu\nu}$ and
 $\overline{R}  = \overline{g}^{\mu \nu}  \overline{R}_{\mu\nu}$ )
\begin{equation}\label{e35}
\overline{R}_{\mu\nu} (\overline{g}_{\alpha\beta}) - \frac{1}{2}
\overline{g}_{\mu\nu}
\overline{R}(\overline{g}_{\alpha\beta}) = \frac{\kappa}{2} T^{eff}_{\mu\nu}
(\phi)
\end{equation}
where
\begin{equation}\label{e36}
T^{eff}_{\mu\nu} (\phi) = \phi_{,\mu} \phi_{,\nu} - \frac{1}{2} \overline
{g}_{\mu\nu} \phi_{,\alpha} \phi_{,\beta} \overline{g}^{\alpha\beta}
+ \overline{g}_{\mu\nu} V_{eff} (\phi)
\end{equation}

and
\begin{equation}\label{e37}
V_{eff} (\phi) = \frac{1}{4U(\phi)}  (V+M)^{2}.
\end{equation}

    In terms of the metric $\overline{g}^{\alpha\beta}$ , the equation
of motion of the Scalar
field $\phi$ takes the standard General - Relativity form
\begin{equation}\label{e38}
\frac{1}{\sqrt{-\overline{g}}} \partial_{\mu} (\overline{g}^{\mu\nu}
\sqrt{-\overline{g}} \partial_{\nu}
\phi) + V^{\prime}_{eff} (\phi) = 0.
\end{equation}

    Notice that if  $V + M = 0,  V_{eff}  = 0$ and $V^{\prime}_{eff}
= 0$ also, provided $V^{\prime}$ is finite and $U \neq 0$ there.
This means the zero cosmological constant state is achieved without
any sort of fine tuning. That is, independently of whether we add to
$V$ a constant piece, or whether we change the value of $M$, as long
as there is still a point where $V+M =0$, then still $ V_{eff}  = 0$
and $V^{\prime}_{eff} = 0$ ( still provided $V^{\prime}$ is finite
and $U \neq 0$ there). This is the basic feature that characterizes
the TMT and allows it to solve the ``old" cosmological constant
problem, at least at the classical level.

In what follows we will study the effective potential (\ref{e37})
for the special case of global scale invariance, which as we will
see displays additional very special features which makes it
attractive in the context of cosmology.

Notice that in terms of the variables $\phi$,
$\overline{g}_{\mu\nu}$, the ``scale" transformation becomes only a
shift in the scalar field $\phi$, since $\overline{g}_{\mu\nu}$ is
invariant (since $\chi \rightarrow \lambda^{-1} \chi$  and
$g_{\mu\nu} \rightarrow \lambda g_{\mu\nu}$)
\begin{equation}\label{e39}
\overline{g}_{\mu\nu} \rightarrow \overline{g}_{\mu\nu}, \phi \rightarrow
\phi - \frac{\theta}{\alpha}.
\end{equation}

If $V(\phi) = f_{1} e^{\alpha\phi}$  and  $U(\phi) = f_{2}
e^{2\alpha\phi}$ as required by scale invariance Eqs. (\ref{e18},
\ref{e20}, \ref{e21}, \ref{e22}, \ref{e23}), we obtain from the
expression (\ref{e37})
\begin{equation}\label{e40}
    V_{eff}  = \frac{1}{4f_{2}}  (f_{1}  +  M e^{-\alpha\phi})^{2}
\end{equation}

    Since we can always perform the transformation $\phi \rightarrow
- \phi$ we can
choose by convention $\alpha > 0$. We then see that as $\phi \rightarrow
\infty, V_{eff} \rightarrow \frac{f_{1}^{2}}{4f_{2}} =$ const.
providing an infinite flat region as depicted in Fig. 1. Also a minimum is achieved at zero
cosmological constant for the case $\frac{f_{1}}{M} < 0$ at the point
\begin{equation}\label{e41}
\phi_{min}  =  \frac{-1}{\alpha} ln \mid\frac{f_1}{M}\mid.
\end{equation}

    In conclusion, the scale invariance of the original theory is
responsible for the non appearance (in the physics) of a certain scale,
that associated to M. However, masses do appear, since the coupling to two
different measures of $L_{1}$ and $L_{2}$ allow us to introduce two
independent
couplings  $f_{1}$ and $f_{2}$, a situation which is  unlike the
standard
formulation of globally scale invariant theories, where usually no stable
vacuum state exists.

The constant of integration $M$ plays a very important role indeed:
any non vanishing value for this constant implements, already at the
classical level S.S.B. of scale invariance.

\section{Generation of two flat regions after the introduction of a $R^{2}$ term}

As we have seen, it is possible to obtain a model that through a spontaneous breaking of scale invariace
can give us a flat region. We want to obtain now two flat regions in our effective potential.
A simple generalization of the action $S_{L}$ will fix this. The basic new feature we add is the presence is higher curvature terms in the action \cite{barrow1}-\cite{mijic3}, which have been shown to be very relevant in cosmology.
In  particular he first inflationary model from
a model with higher terms in the curvature was proposed in \cite{mijic3}.

What one needs to do is simply consider  the
addition of a scale invariant term of the form

\begin{equation}\label{e45}
S_{R^{2}} = \epsilon  \int (g^{\mu\nu} R_{\mu\nu} (\Gamma))^{2} \sqrt{-g} d^{4} x
\end{equation}

The total action being then $S = S_{L} + S_{R^{2}}$.
In the first order formalism $ S_{R^{2}}$ is not only globally scale invariant
but also locally scale invariant, that is conformally invariant (recall that
in the first order formalism the connection is an independent degree of freedom
and it does not transform under a conformal transformation of the metric).

Let us see what the equations of motion tell us, now with the
addition of $S_{R^{2}}$ to the action. First of all, since the
addition has been only to the part of the action that couples to $
\sqrt{-g}$, the equations of motion derived from the variation of
the measure fields remains unchanged. That is Eq.(\ref{e25}) remains
valid.

The variation of the action with respect to $ g^{\mu \nu}$ gives now

\begin{equation}\label{e46}
 R_{\mu\nu} (\Gamma) ( \frac{-\Phi}{\kappa} + 2 \epsilon R  \sqrt{-g}) +
\Phi \frac{1}{2} \phi,_{\mu} \phi,_{\nu} -
\frac{1}{2}(\epsilon R^{2} + U(\phi)) \sqrt{-g} g_{\mu\nu} = 0
\end{equation}

It is interesting to notice that if we contract this equation with
 $ g^{\mu \nu}$ , the $\epsilon$ terms do not contribute. This means
that the same value for the scalar curvature $R$ is obtained as in section 2,
 if we express our result in terms of $\phi$, its derivatives and
$ g^{\mu \nu}$ .
Solving the scalar curvature from this and inserting in the other
$\epsilon$ - independent equation $L_{2} = M$  we get still the same
solution for the ratio of the measures which was found in the case where
the $\epsilon$ terms were absent,
i.e. $\chi =  \frac{\Phi}{\sqrt{-g}}  = \frac{2U(\phi)}{M+V(\phi)}$.

In the presence of the $\epsilon R^{2} $ term in the action, Eq.
(\ref{e26}) gets modified so that instead of $\Phi$,  $\Omega$  =
$\Phi - 2 \epsilon R \sqrt{-g}$ appears. This in turn implies that
Eq.(\ref{e27}) keeps its form but where $\sigma$ is replaced by
$\omega  = ln (\frac{\Omega}{\sqrt{-g}}) =
 ln ( \chi -2\kappa \epsilon R)$,
where once again,
$\chi =  \frac{\Phi}{\sqrt{-g}} = \frac{2U(\phi)}{M+V(\phi)}$.

Following then the same steps as in the model without the curvature square terms, we can then verify that the
connection is the Christoffel symbol of the metric $\overline{g}_{\mu\nu}$
given by

\begin{equation}\label{e47}
\overline{g}_{\mu\nu}   = (\frac{\Omega}{\sqrt{-g}}) g_{\mu\nu}
 = (\chi -2\kappa \epsilon R) g_{\mu\nu}
\end{equation}

$\overline{g}_{\mu\nu} $ defines now the ``Einstein frame".
Equations (\ref{e46}) can now be expressed in the ``Einstein form"

\begin{equation}\label{e48}
\overline{R}_{\mu\nu} -  \frac{1}{2}\overline{g}_{\mu\ \nu} \overline{R} =
\frac{\kappa}{2} T^{eff}_{\mu\nu}
\end{equation}

where

\begin{equation}
 T^{eff}_{\mu\nu} =\label{e49}
\frac{\chi}{\chi -2 \kappa \epsilon R} (\phi_{,\mu} \phi_{,\nu} - \frac{1}{2} \overline
{g}_{\mu\nu} \phi_{,\alpha} \phi_{,\beta} \overline{g}^{\alpha\beta})
+ \overline{g}_{\mu\nu} V_{eff}
\end{equation}

where

\begin{equation}\label{e50}
 V_{eff}  = \frac{\epsilon R^{2} + U}{(\chi -2 \kappa \epsilon R)^{2} }
\end{equation}

Here it is satisfied that $\frac{-1}{\kappa} R(\Gamma,g) +
\frac{1}{2} g^{\mu\nu}\partial_{\mu} \phi \partial_{\nu} \phi - V = M $,
equation that expressed in terms of $ \overline{g}^{\alpha\beta}$
 becomes

$\frac{-1}{\kappa} R(\Gamma,g) + (\chi -2\kappa \epsilon R)
\frac{1}{2} \overline{g}^{\mu\nu}\partial_{\mu} \phi \partial_{\nu} \phi - V = M$.
 This allows us to solve for $R$ and we get,

\begin{equation}\label{e51}
R = \frac{-\kappa (V+M) +\frac{\kappa}{2} \overline{g}^{\mu\nu}\partial_{\mu} \phi \partial_{\nu} \phi \chi}
{1 + \kappa ^{2} \epsilon \overline{g}^{\mu\nu}\partial_{\mu} \phi \partial_{\nu} \phi}
\end{equation}

Notice that
 if we express $R$ in
terms of $\phi$, its derivatives and $ g^{\mu \nu}$, the result is the
same as in the model without the curvature squared term, this is not true anymore once we express
 $R$ in terms of $\phi$, its derivatives and $\overline{g}^{\mu\nu}$.

In any case, once we insert (\ref{e51}) into (\ref{e50}), we see
that the  effective potential  (\ref{e50}) will depend on the
derivatives of the scalar field now. It acts as a normal scalar
field potential under the conditions of slow rolling  or low
gradients and in the case the scalar field is near the region
$M+V(\phi ) = 0$.

Notice that since
$\chi =   \frac{2U(\phi )}{M+V(\phi )}$,
then if ${M+V(\phi) = 0}$, then, as in the simpler model without the curvature squared terms, we obtain that
 $ V_{eff}  =  V'_{eff} =  0$ at that point without fine tuning
(here by $ V'_{eff}$ we mean the derivative  of $ V_{eff}$ with
respect to the scalar field $\phi$, as usual).

In the case of the scale invariant case, where $V$ and $U$ are given
by equation (\ref{e19}), it is interesting to study the shape of $
V_{eff} $ as a function of $\phi$ in the case of a constant $\phi$,
in which case $ V_{eff} $ can be regarded as a real scalar field
potential. Then from (\ref{e51}) we get $R = -\kappa (V+M)$, which
inserted in (\ref{e50}) gives,
\begin{equation}\label{effpotslow}
 V_{eff}  =
\frac{(f_{1} e^{ \alpha \phi }  +  M )^{2}}{4(\epsilon \kappa ^{2}(f_{1}e^{\alpha \phi}  +  M )^{2} + f_{2}e^{2 \alpha \phi })}
\end{equation}

The limiting values of $ V_{eff} $ are:

First, for asymptotically
large positive values, ie. as $ \alpha\phi \rightarrow  \infty $,
we have
$V_{eff} \rightarrow
\frac{f_{1}^{2}}{4(\epsilon \kappa ^{2} f_{1}^{2} + f_{2})} $.

Second, for asymptotically large but negative values of the scalar field,
that is as $\alpha \phi \rightarrow - \infty  $ ,  we have:
$ V_{eff} \rightarrow \frac{1}{4\epsilon \kappa ^{2}}$ .

In these two asymptotic regions ($\alpha \phi \rightarrow  \infty  $
and $\alpha \phi \rightarrow - \infty  $) an examination of the scalar
field equation reveals that a constant scalar field configuration is a
solution of the equations, as is of course expected from the flatness of
the effective potential in these regions.

Notice that in all the above discussion it is fundamental that $ M\neq 0$.
If $M = 0$ the potential becomes just a flat one,
$V_{eff} = \frac{f_{1}^{2}}{4(\epsilon \kappa ^{2} f_{1}^{2} + f_{2})}$
everywhere (not only at high values  of $\alpha \phi$). All the non trivial
features necessary for a graceful exit, the other flat
region associated to the Planck scale and the minimum at zero if $M<0$ are all lost .
As we discussed in the model without a curvature squared term, $ M\neq 0$ implies the we are considering a
situation with S.S.B. of scale invariance.

These kind of models with potentials giving rise to two flat
potentials have been applied to produce models for bags and
confinement in a very natural way \cite{bags and confinement}.

\section{A Note on the the ``Einstein" metric as a canonical variable of the Theory}
One could question the use of the Einstein frame metric
$\overline{g}_{\mu\nu} $ in contrast to the original metric
$g_{\mu\nu} $. In this respect, it is interesting to see the role of
both the original metric and that of the Einstein frame metric in a
canonical approach to the first order formalism. Here we see that
the original metric does not have a canonically conjugated momentum
(this turns out to be zero), in contrast, the canonically conjugated
momentum to the connection turns out to be a function exclusively of
$\overline{g}_{\mu\nu}$, this Einstein metric is therefore a genuine
dynamical canonical variable, as opposed to the original metric.
There is also a lagrangian formulation of the theory which uses
$\overline{g}_{\mu\nu}$, as we will see in the next section, what we
can call the action in the Einstein frame. In this frame we can
quantize the theory for example and consider contributions without
reference to the original frame, thus possibly considering breaking
the TMT structure of the theory through quantum effects, but such
breaking will be done ``softly" through the introduction of a
cosmological term only. Surprisingly, the remaining structure of the
theory, reminiscent from the original TMT structure will be enough
to control the strength of this additional cosmological term once we
demand that the universe originated from a non-singular and stable
emergent state.

\section{Generalizing the model to include effects of zero point fluctuations}

The effective energy-momentum tensor can be
represented in a form like that of  a perfect fluid
\begin{equation}
T_{\mu\nu}^{eff}=(\rho +p)u_{\mu}u_{\nu}-p\tilde{g}_{\mu\nu},
\qquad \text{where} \qquad
u_{\mu}=\frac{\phi_{,\mu}}{(2X)^{1/2}}\label{Tmnfluid}
\end{equation}
here $X\equiv\frac{1}{2}\tilde{g}^{\alpha\beta}\phi_{,\alpha}\phi_{,\beta}$. This defines a pressure functional and an energy density functional.
The system of equations obtained after solving for $\chi$, working
in the Einstein frame with the metric
$\tilde{g}_{\mu \nu}$ can be obtained from a
 ``k-essence" type effective action, as it is standard in treatments
 of theories with non linear kinetic terms or k-essence models\cite{k-essence1}-\cite{k-essence4}. The action from which the classical equations follow is,
\begin{equation}
S_{eff}=\int\sqrt{-\overline{g}}d^{4}x\left[-\frac{1}{\kappa}\overline{R}(\overline{g})
+p\left(\phi,R\right)\right] \label{k-eff}
\end{equation}

\begin{equation}
 p = \frac{\chi}{\chi -2 \kappa \epsilon R}X - V_{eff}
\end{equation}

\begin{equation}\label{e500}
 V_{eff}  = \frac{\epsilon R^{2} + U}{(\chi -2 \kappa \epsilon R)^{2} }
\end{equation}

where it is understood that,
\begin{equation}\label{chi}
\chi = \frac{2U(\phi)}{M+V(\phi)}.
\end{equation}
We have two possible formulations concerning $R$:
Notice first that $\overline{R}$ and $R$ are different objects, the $\overline{R}$ is the Riemannian curvature scalar in the Einstein frame,
while $R$ is a different object. This $R$ will be treated in two different ways:

1. First order formalism for $R$. Here $R$ is a lagrangian variable,
determined as follows,  $R$ that appear in the expression above for
$p$ can be obtained from the variation of the pressure functional
action above with respect to $R$, this gives exactly the expression
for $R$ that has been solved already in terms of $X, \phi$, etc, see
Eq. (\ref{e51}).

2. Second order formalism for $R$. $R$ that appear in the action
above is exactly the expression for $R$ that has been solved already
in terms of $X, \phi$, etc. The second order formalism can be
obtained from the first order formalism by solving algebraically R
from the Eq. (\ref{e51}) obtained by variation of $R$ , and
inserting back into the action.

One may also use the method outlined in \citep{R2Bulg} to find the effective action in the Einstein frame, in \citep{R2Bulg} the problem of a curvature squared theory with standard measure was studied. The methods outlined there can be also applied in the modified measure case \citep{mahary}, thus providing another derivation of the effective action explained above.

The problem that we have to solve to find the effective lagrangian is basically finding that lagrangian tat will produce the effective energy momentum tensor in the Einstein frame by the variation of
the $\overline{g}_{\mu\nu}$ metric
\begin{equation}
{T_{\rm eff}}_{\mu\nu}  = \overline{g}_{\mu\nu} L_{\rm eff} (h) -
2 \partder{L_{\rm eff}}{\overline{g}^{\mu\nu}}
\lab{T-h-eff}
\end{equation}


\begin{figure}[t]
\centering
\includegraphics[width=7cm]{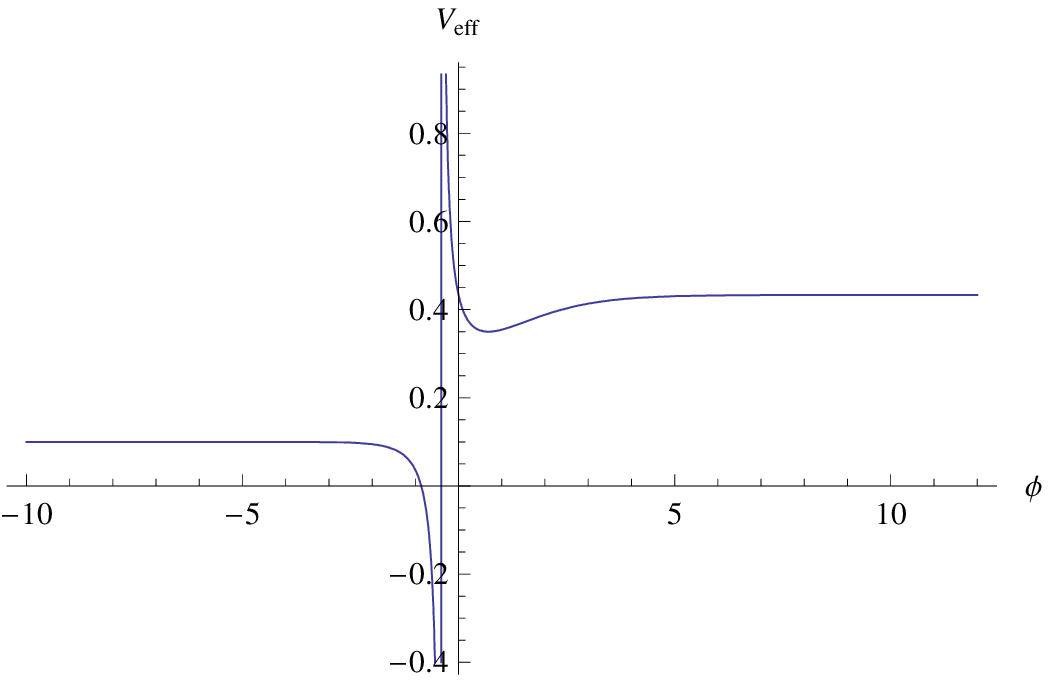}
\includegraphics[width=7cm]{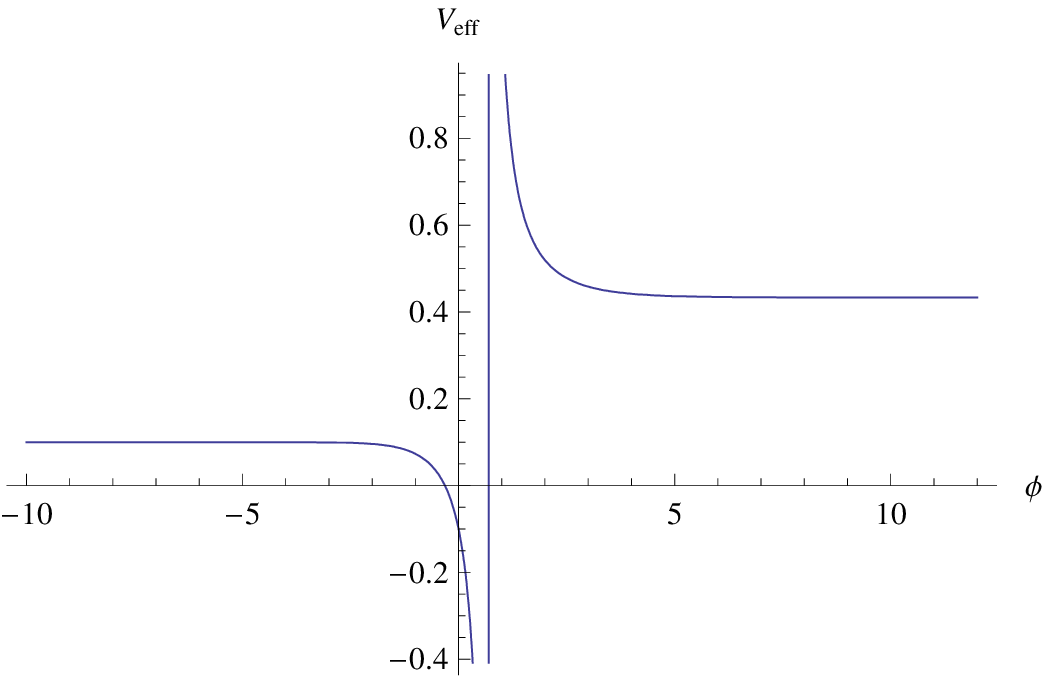}
\caption{The form of effective potential $V_{eff}(\phi)$ versus the
scalar field $\phi$. We consider unit where $\kappa = 1$,
$\alpha=1$, $\Lambda = 0.35$ and $\epsilon = -1$. Left panel: $M
=-1$, $f_1=1/2$, $f_2=1$. Right panel: $M =1$, $f_1=1/2$, $f_2=1$.
\label{Fig-Paper1}}
\end{figure}


In contrast to the simplified models studied in
literature\cite{k-essence1, k-essence2, k-essence3, k-essence4}, it
is impossible here to represent $p\left(\phi,X;M\right)$ in a
factorizable form like $\tilde{K}(\phi)\tilde{p}(X)$. The scalar
field effective Lagrangian can be taken as a starting point for many
considerations.

In particular, the quantization of the model can proceed from
(\ref{k-eff}) and additional terms could be generated by radiative
corrections. We will focus only on a possible cosmological term in
the Einstein frame added (due to zero point fluctuations) to
(\ref{k-eff}), which leads then to the new action
\begin{equation}
S_{eff,\Lambda }=\int\sqrt{-\overline{g}}d^{4}x\left[-\frac{1}{\kappa}\overline{R}(\overline{g})
+p\left(\phi,R\right)- \Lambda \right] \label{act.lambda}
\end{equation}

This addition to the effective action leaves the equations of motion of the scalar field unaffected, but the gravitational equations aquire a
cosmological constant. Adding the $\Lambda$ term can be regarded as a redefinition of $V_{eff}\left(\phi,X;M\right)$
\begin{equation}
V_{eff}\left(\phi,R\right) \rightarrow V_{eff}\left(\phi,R\right) + \Lambda  \label{V.lambda}
\end{equation}As we will see the stability of the emerging Universe imposes interesting constraints on $\Lambda$

\begin{figure}[t]
\centering
\includegraphics[width=7cm]{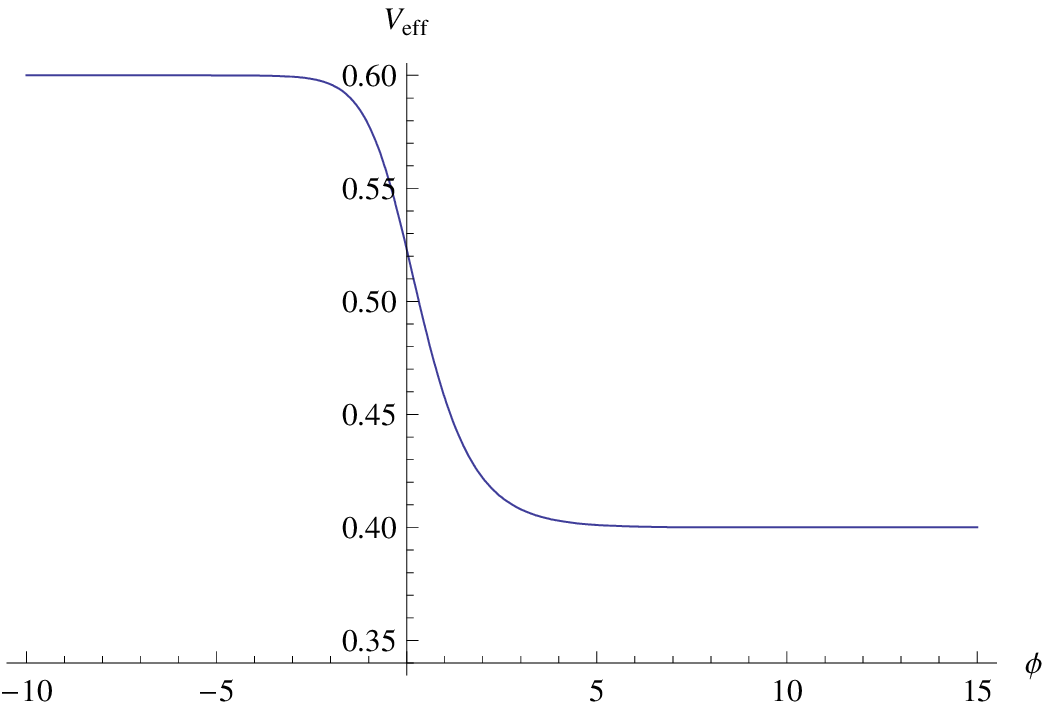}
\includegraphics[width=7cm]{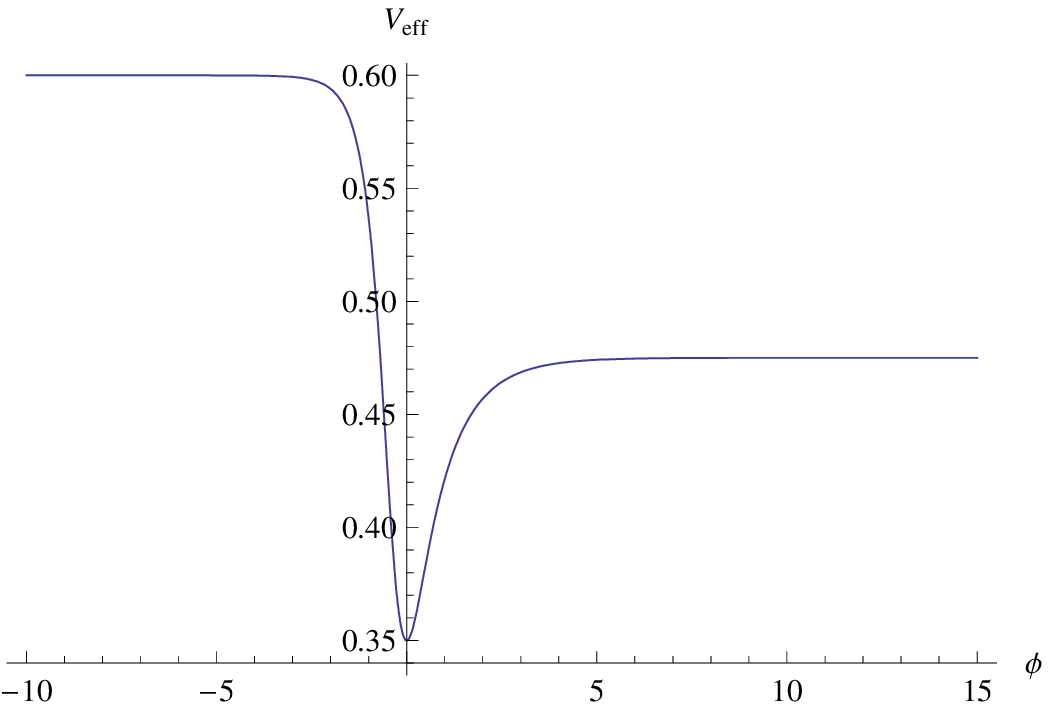}
\caption{The effective potential $V_{eff}(\phi)$ versus the scalar
field $\phi$. We consider unit where $\kappa = 1$, $\alpha=1$,
$\Lambda = 0.35$ and $\epsilon = 1$. Left panel: $M =1$, $f_1=1/2$,
$f_2=1$. Right panel: $M =-1$, $f_1=1$, $f_2=1$.\label{Fig-Paper2}}
\end{figure}

After introducing the $\Lambda$ term, we get from the variation of
$R$ the same value of $R$, unaffected by the new $\Lambda$ term, but
as one can easily see then $R$ does not have the interpretation of a
curvature scalar in the original frame since it is unaffected by the
new source of energy density (the $\Lambda$ term), this is why the
$\Lambda$ term theory does not have a formulation in the original
frame, but is a perfectly legitimate generalization of the theory,
probably obtained by considering zero point fluctuations, notice
that quantum theory is possible only in the Einstein frame. Notice
that even in the original frame  the bar metric (not the original
metric) appears automatically in the canonically conjugate momenta
to the connection, so we can expect from this that the bar metric
and not the original metric be the relevant one for the quantum
theory.


In Figure \ref{Fig-Paper1} and \ref{Fig-Paper2} we have plotted the
effective potential as a function of the scalar field, for $\epsilon
= -1$ and $\epsilon = 1$ respectively. We consider unit where
$\kappa =1$, $\alpha=1$, $\Lambda = 0.35$ and different values for
$M,f_1, f_2$.

\section{Analysis of the Emergent Universe solutions}

We now want to consider the detailed analysis of The Emerging
Universe solutions and in the next section their stability in the
TMT scale invariant theory. We start considering the cosmological
solutions of the form
\begin{equation}
ds^2 =dt^2 - a(t)^2 (\frac{dr^2}{1 -r^2}+ r^2(d\theta^2 +sin^2\theta d\phi^2)),   \phi = \phi(t)
\end{equation}

in this case, we obtain for the energy density and the pressure, the following expressions.
We will consider a scenario where the scalar field $\phi$ is moving in the extreme right region
$\phi \rightarrow \infty  $, in this case the expressions for the energy density $\rho$ and pressure $p$ are given by,
\begin{equation}\label{eq.density}
\rho = \frac{A}{2} \dot{\phi}^2 + 3B\dot{\phi}^4 + C
\end{equation}

and
\begin{equation}
p = \frac{A}{2} \dot{\phi}^2 +B\dot{\phi}^4 - C
\end{equation}
It is interesting to notice that all terms proportional to
$\dot{\phi}^4$ behave like ``radiation", since
 $p_{\dot{\phi}^4} = \frac{\rho_{\dot{\phi}^4} }{3}$ is satisfied.
here the constants $A,B$ and $C$ are given by,

\begin{eqnarray}\label{ABC}
A &=& \frac{f_2}{f_2 + \kappa^2\epsilon f_1^2}\,,\\
B &=& \frac{\epsilon\kappa^2}{4(1+\kappa^2\epsilon f_1^2/f_2)} = \frac{\epsilon \kappa^2}{4}\,A \,,\label{B} \\
C &=& \frac{f_1^2}{4\,f_2(1+\kappa^2\epsilon f_1^2/f_2)}  + \Lambda =
\frac{f_1^2}{4f_2}\,A + \Lambda\,\label{C}.
\end{eqnarray}
It will be convenient to ``decompose" the constant $\Lambda$ into
two pieces,

\begin{eqnarray}\label{Lambda}
\Lambda = -\frac{1}{4\kappa^2\epsilon} + \Delta \lambda
\end{eqnarray}
since as $\phi \rightarrow -\infty $ , $ V_{eff} \rightarrow  \Delta
\lambda $. Therefore $\Delta \lambda$ has the interesting
interpretation of the vacuum energy density in the $\phi \rightarrow
-\infty $ vacuum. As we will see, it is remarkable that the
stability and existence of non-singular emergent universe implies
that $\Delta \lambda > 0$, and it is bounded from above as well.

The equation that determines such static universe $a(t) = a_0 =constant$,
$\dot{a}=0$, $\ddot{a}=0$ gives rise to a restriction for $\dot{\phi}_0$
that have to satisfy the following equation in order to
guarantee that the universe be static, because $\ddot{a}=0$ is proportional to
$\rho + 3p$, we must require that $\rho + 3p = 0$, which leads to

\begin{equation}\label{e1}
3B\dot{\phi}^4_0 + A\dot{\phi}^2_0 - C=0,
\end{equation}

This equation leads to two roots, the first being

\begin{equation}\label{e2}
\dot{\phi}_1^2=\frac{\sqrt{A^2+ 12BC}\,-A}{6B}\,.
\end{equation}

The second root is:

\begin{equation}\label{e3}
\dot{\phi}_2^2=\frac{-\sqrt{A^2+ 12BC}\,-A}{6B}\,.
\end{equation}

It is also interesting to see that if the discriminant is positive, the first solution has automatically positive energy density, if we only consider cases where $C>0$, which is required if we want the emerging solution to be able to turn into an inflationary solution eventually. One can see that the condition $\rho >0$ for the first solution reduces to the inequality $w> (1-\sqrt{1-w} )/2$, where $w =-12BC/A^2 >0$, since we must have $A>0$, otherwise we get a negative kinetic term during the inflationary period, and as we will see in the next section, we must have that $B<0$ from the stability of the solution, and as long as $w<1$, it is always true that this inequality is satisfied.

Before going into the subject of the small perturbations and
stability of these solutions, we would like to notice the ``entropy
like" conservation laws that may be useful in a non perturbative
analysis of the theory.

In fact in the $\phi \rightarrow \infty$ region, we have the exact symmetry $\phi \rightarrow \phi + constant$. and considering that the effective matter action here is $a^3 p$, we have the conserved quantity

\begin{equation}
\pi_\phi = a^3( A \dot{\phi}+4B\dot{\phi}^3)
\end{equation}

It is very interesting to notice that
\begin{equation}
\pi_\phi = S= a^3 s
\end{equation}
where $s$ assumes the ``entropy density" form
\begin{equation}
s =(\rho + p)/ T
\end{equation}
provided we identify the ``Temperature" T with $\dot{\phi}$.

\section{Stability of the static solution}

We will now consider the
perturbation equations. Considering small deviations of $\dot{\phi}$ the from the static emerging solution
value $\dot{\phi}_0$ and also considering the perturbations of the scale factor $a$, we obtain, from
Eq.~(\ref{eq.density})

\begin{equation}\label{eq.density-pert.}
\delta \rho = A \dot{\phi}_0 \delta \dot{\phi} + 12B \dot{\phi}_0^3 \delta \dot{\phi}
\end{equation}

at the same time $\delta \rho$ can be obtained from the perturbation of the Friedmann equation

\begin{equation}\label{Fried.eq.}
3(\frac{1}{a^2}+H^2)=\kappa \rho
\end{equation}
and since we are perturbing a solution which is static, i.e., has $H=0$, we obtain then
\begin{equation}\label{pert.Fried.eq.}
-\frac{6}{a_0^3}\delta a =\kappa \delta \rho
\end{equation}

we also have the second order Friedmann equation

\begin{equation}\label{Fried.eq.2}
\frac{1+\dot{a}^2 + 2a\ddot{a}}{a^2}=-\kappa p
\end{equation}

For the static emerging solution, we have $p_0=-\rho_0/3$, $a=a_0$,  so
\begin{equation}
\frac{2}{a_0^2} = -2\kappa p_0 = \frac{2}{3}\kappa \rho_0= \Omega_0 \kappa \rho_0
\end{equation}
where we have chosen to express our result in terms of $\Omega_0$, defined by $p_0=(\Omega_0-1)\rho_0$, which for the emerging
solution has the value $\Omega_0=\frac{2}{3}$. Using this in \ref{pert.Fried.eq.}, we obtain
\begin{equation}\label{pert.Fried.eq.3}
\delta \rho = -\frac{3\Omega_0 \rho_0}{a_0}\delta a
\end{equation}
and equating the values of $\delta \rho$ as given by \ref{eq.density-pert.} and \ref{pert.Fried.eq.3} we obtain a linear relation between
$\delta \dot{\phi}$ and $\delta a$, which is,
\begin{equation}\label{delta-delta}
\delta \dot{\phi}=D_0\delta a
\end{equation}
where

\begin{equation}
D_0 = -\frac{3\Omega_0 \rho_0}{a_0 \dot{\phi}_0 (A + 12 B \dot{\phi}_0^2)}
\end{equation}

we now consider the perturbation of the eq. (\ref{Fried.eq.2}). In
the right hand side of this equation we consider that
$p=(\Omega-1)\rho$, with
\begin{equation}\label{Omega-eq.}
\Omega = 2\Big(1 - \frac{U_{eff}}{\rho}\Big),
\end{equation}
where,
\begin{equation}\label{V-eq.}
U_{eff} =C + B\,\dot{\phi}^4
\end{equation}

and therefore,  the perturbation of the Eq. (\ref{Fried.eq.2}) leads
to,

\begin{equation}\label{pert.Fried.eq.2}
-\frac{2\delta a}{a_0^3}+2\frac{\delta\ddot{a}}{a_0}=-\kappa \delta p =-\kappa \delta ((\Omega-1)\rho)
\end{equation}

to evaluate this, we use \ref{Omega-eq.}, \ref{V-eq.} and the
expressions that relate the variations in $a$ and $\dot{\phi}$
(\ref{delta-delta}).  Defining the ``small"  variable $\beta$ as

\begin{equation}
a(t) = a_0( 1+ \beta)
\end{equation}
we obtain,
\begin{equation}
2\ddot{\beta}(t) + W_0^2\beta(t) = 0\,,
\end{equation}

where,
\begin{equation}
W_0^2 = \Omega_0\,\rho_0\left[ \frac{24\,B\,\dot{\phi}_0^2}{A +
12\,\dot{\phi}_0^2\,B }  -6\frac{(C + B\,
\dot{\phi}_0^4)}{\rho_0} -3\kappa \Omega_0 + 2\kappa \right],
\end{equation}

notice that the sum of the last two terms in the expression for $W_0^2$, that is $-3\kappa \Omega_0 + 2\kappa $  vanish
since $\Omega_0=\frac{2}{3}$, for the same reason, we have that $6\frac{(C + B\,\dot{\phi}_0^4)}{\rho_0} = 4$, which brings us to the simplified expression
\begin{equation}
W_0^2 = \Omega_0\,\rho_0\left[ \frac{24\,B\,\dot{\phi}_0^2}{A +
12\,\dot{\phi}_0^2\,B }  - 4 \right],
\end{equation}

For the stability of the static solution, we need that  $W_0^2 >0$,
where $\dot{\phi}_0^2$ is defined either by E.~(\ref{e2}) ($\dot{\phi}_0^2=\phi_1^2$) or by E.~(\ref{e3}) ($\dot{\phi}_0^2=\phi_2^2$).
If we take E.~(\ref{e3}) ($\dot{\phi}_0^2=\phi_2^2$) and use this in the above expression for $W_0^2$, we obtain,
\begin{equation}
W_0^2 = \Omega_0\,\rho_0\left[ \frac{4\sqrt{A^2 +12BC}}{-2\sqrt{A^2 +12BC} -A}  \right],
\end{equation}

to avoid negative kinetic terms during the slow roll phase that takes place following the emergent phase, we must consider $A>0$, so, we  see that the second solution is unstable and will not be considered further.

Now in the case of the first solution, E.~(\ref{e2}) ($\dot{\phi}_0^2=\phi_1^2$), then $W_0^2$ becomes
\begin{equation}
W_0^2 = \Omega_0\,\rho_0\left[ \frac{-4\sqrt{A^2 +12BC}}{2\sqrt{A^2 +12BC} -A}  \right],
\end{equation}
so the condition of stability becomes $2\sqrt{A^2 +12BC} -A < 0$, or $2\sqrt{A^2 +12BC} < A  $, squaring both sides and since $A>0$, we get
$12BC/A^2 < -3/4$, which means $B<0$, and therefore $\epsilon <0$,  multiplying by $-1$, we obtain, $12(-B)C/A^2 > 3/4$, replacing the values of $A, B, C$, given by \ref{ABC} we obtain the condition

\begin{equation}
 \Delta\lambda > 0,
\end{equation}

Now there is the condition that the discriminant be positive $A^2 +12BC> 0$

\begin{equation}
 \Delta\lambda < \frac{1}{12(-\epsilon)\kappa^2 } \left[ \frac{f_2}{f_2 + \kappa^2 \epsilon f^2_1 }  \right],
\end{equation}

since $A=\left[ \frac{f_2}{f_2 + \kappa^2 \epsilon f^2_1 }  \right]> 0$, $B<0$, meaning that $\epsilon < 0$, we see that we obtain a positive upper bound for the energy density of the vacuum as $\phi \rightarrow -\infty $, which must be positive, but not very big.

 \section{Inflation and its Graceful Exit}
The emerging phase owes its existence to a strictly constant vacuum
energy (which here is represented by the value of $A$) at very large
values of the field $\phi$. In fact, while for $M=0$ the effective
potential of the scalar field is perfectly flat, for any $M \neq 0$
the effective potential acquires a non trivial shape. This causes the transition from
the emergent phase to a slow roll inflationary phase which will be the subject
of this section.

Following  \cite{SIchile}, we consider now then the relevant equations for the model in the slow
roll regime, i.e. for $\dot{\phi}$ small and when the scalar field
$\phi$ is large, but finite and we consider the first corrections to
the flatness to the effective potential. Dropping higher powers of
$\dot{\phi}$ in the contributions for the kinetic energy and in the
scalar curvature $R$, we obtain
\begin{eqnarray}
\rho =  \frac{1}{2}\gamma\dot{\phi}^2 + V_{eff}, \label{energydensityslow}\\
\gamma = \frac{\chi}{\chi -2\kappa\epsilon R}, \label{gamma}\\
R = -\kappa (V+M)\label{R}.
\end{eqnarray}
Here, as usual  $\chi= \frac{2U(\phi)}{M+V(\phi)}$. In the slow roll
approximation, we can drop the second derivative term of $\phi$ and
the second power of $\dot{\phi}$ in the equation for $H^2$ and we
get
\begin{eqnarray}
3H \gamma \dot{\phi} = - V^{\prime}_{eff} \label{slowroll},\\
3H^2 = \kappa V_{eff} \label{Friedmannslow},
\end{eqnarray}
where $ V^{\prime}_{eff}= \frac{dV_{eff}}{d\phi}$. The relevant
expression for $V_{eff}$ will be that given by (\ref{effpotslow}),
i.e., where all higher derivatives are ignored in the potential,
consistent with the slow roll approximation.

We now display the relevant expressions for the region of very
large, but not infinite $\phi$, these are:
\begin{eqnarray}
 V_{eff} = C + C_1 exp(-\alpha \phi),\label{effpotatlargephi}\\
\chi =2 \frac{f_2}{f_1}exp(\alpha \phi)-
2M\frac{f_2}{f^2_1},\label{chiatlargeph}
\end{eqnarray}
and
\begin{eqnarray} \gamma = \gamma_0 + \gamma_1 exp(-\alpha
\phi). \label{gammaatlargephi}
\end{eqnarray}

The relevant constants that will affect our results are, $C$, as
given by (\ref{C}) and $C_1 $ and $ \gamma_0$ given by
\begin{eqnarray}
  C_1 = -\frac{8\epsilon \kappa^2 f^3_1 M}{(4f_2+4\kappa^2\epsilon f_1^2)^2}+
  \frac{2 f_1 M}{4f_2+4\kappa^2\epsilon f_1^2},\label{C1}\end{eqnarray}
and
  \begin{eqnarray}
 \gamma_0 = \frac{ f_2}{f_2+\kappa^2\epsilon f_1^2}, \label{gamma0}
\end{eqnarray}
respectively.

Using Eq. (\ref{effpotatlargephi}) we can calculates the key
landmarks of the inflationary history: first, the value of the
scalar field where inflation ends, $\phi_{end}$ and a value for the
scalar field $\phi_{*}$ bigger than this ($\phi_{*} > \phi_{end}$)
and which happens earlier, which represents the ``horizon crossing
point". We must demand then that a typical number of e-foldings,
like $N=60$, takes place between $\phi_{*}$, until the end of
inflation at $\phi=\phi_{end}$.

To determine the end of inflation, we consider the quantity $\delta
= - \frac{\dot{H}}{H^2}$ and consider the point in the evolution of
the Universe where $\delta =1$, only when $\delta < 1$, we have an
accelerating Universe, so the point $\delta =1$ represents indeed
the end of inflation. Calculating the derivative with respect to
cosmic time of the Hubble expansion using (\ref{Friedmannslow}) and
(\ref{slowroll}), we obtain that the condition $\delta =1$ gives

\begin{equation}
 \delta =\frac{1}{2\gamma}(V^{\prime}_{eff}/V_{eff})^2 = 1, \label{endofinf.eq.}
\end{equation}
 working to leading order, setting $\gamma = \gamma_0$, $V_{eff} = C $
 and $V^{\prime}_{eff}= - \alpha C_1 exp(-\alpha \phi_{end})$, this gives as a solution,
\begin{equation}
 exp(\alpha \phi_{end})= -\frac{\alpha C_1 }{C \sqrt{2\kappa
 \gamma_0}},
 \label{sol.endofinf.eq.}
\end{equation}
notice that if $M$ and $f_1$ have different signs and if
$\epsilon<0$, $C_1<0$ for the allowed range of  parameters the
stable emerging solution, so $-C_1$ represents the absolute value of
$C_1$. We now consider $\phi_{*}$ and the requirement that this
precedes $\phi_{end}$ by $N$ e-foldings,
\begin{equation}
N= \int Hdt = \int \frac{H}{\dot{\phi}} d\phi = -\int
\frac{3H^2\gamma}{ V^{\prime}_{eff}} d\phi, \label{N}
\end{equation}
where in the last step we have used the slow roll equation of motion
for the scalar field (\ref{slowroll}) to solve for $\dot{\phi}$.
Solving $H^2$ in terms of $V_{eff}$ using (\ref{Friedmannslow}),
working to leading order, setting $\gamma = \gamma_0$ and
integrating, we obtain the relation between $\phi_{*}$ and
$\phi_{end}$,
\begin{equation}\label{sol.crossing.}
exp(\alpha \phi_{*}) = exp(\alpha \phi_{end}) - \frac{N \alpha^2 C_1
}{C\kappa \gamma_0},
\end{equation}
as we mentioned before $C_1<0$ for the allowed range of parameters
the stable emerging solution, so that $\phi_{*} > \phi_{end}$ as it
should be for everything to make sense. Introducing Eq.
(\ref{sol.endofinf.eq.}) into Eq. (\ref{sol.crossing.}), we obtain,
\begin{equation}\label{sol.crossing.final}
exp(\alpha \phi_{*}) = -\frac{ C_1 }{C\sqrt{\kappa}}( \frac{\alpha}
{\sqrt{2\gamma_0}} + \frac{N\alpha^2}{\sqrt{\kappa}\gamma_0} ).
\end{equation}

We finally calculate the power of the primordial scalar
perturbations. If the scalar field $\phi$ had a canonically
normalized kinetic term, the spectrum of the primordial
perturbations will be given by the equation
 \begin{equation}\label{primordialpert.}
 \frac{\delta \rho }{\rho} \propto \frac{H^2}{\dot{\phi}},
 \end{equation}
however, as we can see from (\ref{energydensityslow}), the kinetic
term is not canonically normalized because of the factor $\gamma$ in
that equation.

In this point we will study the scalar and tensor perturbations for
our model where the kinetic term is not canonically normalized. The
general expression for the perturbed metric about the
Friedmann-Robertson-Walker is
\begin{eqnarray*}
ds^2&=&-(1+2 F) dt^2+ 2 a(t) D_{,\,i}dx^i dt + a^2(t) [(1-2 \psi)
\delta_{ij}+2 E_{,i,j}+2 h_{ij}] dx^i dx^j,
\end{eqnarray*}
where  $F$, $D$, $\psi$ and $E$ are the scalar type metric
perturbations and $h_{ij}$ characterizes the transverse-traceless
tensor perturbation. The power spectrum of the curvature
perturbation in the slow-roll approximation for a not-canonically
kinetic term becomes Ref.\cite{Garriga}(see also Refs.\cite{p1})

\begin{equation}
P_S = k_1\left(\frac{\delta \rho
}{\rho}\right)^2=k_1\,\frac{H^2}{c_s\,\delta},\label{pec}
\end{equation}
where it was  defined  ``speed of sound", $c_s$, as
$$
c_s^2=\frac{P_{,\,X}}{P_{,\,X}+2XP_{,\,XX}},
$$
 with $P(X,\phi)$  an  function of the scalar field and of the
 kinetic term $X=-(1/2)g^{\mu\nu}\partial_\mu \phi\partial_\nu \phi$.
 Here $P_{,\,X}$ denote the derivative with respect $X$.
 In our case $P(X,\phi)=\gamma(\phi)\,X-V_{eff}$, with $X=\dot{\phi}^2/2$.
 Thus, from Eq.(\ref{pec}) we get
\begin{equation}
P_S=k_1\frac{H^4}{\gamma(\phi) \dot{\phi}^2}.\label{eq10}
\end{equation}

 The scalar spectral index $n_s$, is defined by
\begin{equation}
n_s-1=\frac{d\ln P_S}{d\ln k}=-2\delta-\eta-s,\label{ns}
\end{equation}
where $\eta=\frac{\dot{\delta}}{\delta\,H}$ and
$s=\frac{\dot{c_s}}{c_s\,H}$, respectively.

On the other hand, the generation of tensor perturbations during
inflation  would produce gravitational wave. The amplitude of tensor
perturbations was evaluated in Ref.\cite{Garriga}, where
$$
P_T=\frac{2}{3\pi^2}\,\left(\frac{2XP_{,\,X}-P}{M_{Planck}^4}\right),
$$
and the tensor spectral index $n_T$, becomes
$$
n_T=\frac{d\ln P_T}{d\ln k}=-2\delta,
$$
and they satisfy a generalized consistency relation
\begin{equation}
r=\frac{P_T}{P_S}=-8\,c_s\,n_T. \label{ration}
\end{equation}

\begin{figure}[th]
\includegraphics[width=5.0in,angle=0,clip=true]{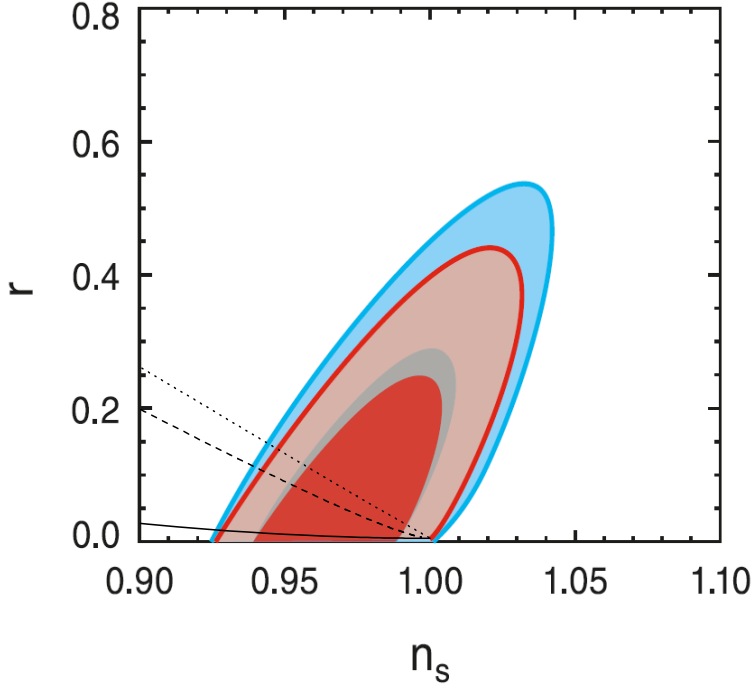}
\caption{ The plot shows $r$ versus $n_s$ for three values of
$\alpha$. For $\alpha=1$ solid line,  $\alpha=0.1$ dash line and
$\alpha=0.01$ dots line, respectively. Here, we have fixed the
values $M=-1$, $\epsilon=-1$, $f_1 = 1/2$, $f_2=1$, $\lambda=1/10$
and $\kappa=1$, respectively. The seven-year WMAP data places
stronger limits on the tensor-scalar ratio (shown in red) than
five-year data (blue) \cite{Larson:2010gs}.\label{fig2}}
\end{figure}

Therefore, the scalar field (to leading order) that should appear in
Eq. (\ref{primordialpert.}) should be $\sqrt{\gamma_0}\phi$ and
instead of Eq. (\ref{eq10}) , we must use
\begin{equation}\label{primordialpert.corr}
 \frac{\delta \rho }{\rho} =
 \frac{H^2}{\sqrt{\gamma_0}\dot{\phi}}.
 \end{equation}
The power spectrum of the perturbations goes, up to a factor of
order one, which we will denote $k_1$ as $(\delta \rho /\rho )^2$,
so we have,
\begin{equation}\label{primordialpert.power}
P_S = k_1\left(\frac{\delta \rho }{\rho}\right)^2 =
k_1\frac{H^4}{\gamma_0 \dot{\phi}^2},
 \end{equation}
this quantity should be evaluated at $\phi=\phi_{*}$ given by
(\ref{sol.crossing.final}). Solving for $\dot{\phi}$ from the slow
roll equation (\ref{slowroll}), evaluating the derivative of the
effective potential using (\ref{effpotatlargephi}) and solving for
$H$ from (\ref{Friedmannslow}), we obtain, to leading order,
\begin{equation}\label{primordialpert.power-almostfinal}
P_S = k_1\frac{\kappa^3 \gamma_0 C^3 }{3\alpha^2 C_1}
exp(2\alpha\phi_{*}),
\end{equation}
using then (\ref{sol.crossing.final}) for $\exp(\alpha\phi_{*})$, we
obtain our final result,
\begin{equation}\label{primordialpert.power-final}
P_S = k_1\frac{\kappa^2 C }{3}(\frac{1}{\sqrt{2}} + \frac{N
\alpha}{\sqrt{\gamma_0 \kappa}})^2,
\end{equation}
it is very interesting first of all that $C_1$ dependence has
dropped out and with it all dependence on $M$. In fact this can be
regarded as a non trivial consistency check of our estimates, since
apart from its sign, the value $M$ should not affect the results.
This is due to the fact that from a different value of $M$ (although
with the same sign), we can recover the original potential by
performing a shift of the scalar field $\phi$.

In Fig.\ref{fig2} we show the dependence of the tensor-scalar ratio
$r$ on the spectral index $n_s$. From left to right $\alpha=1$
(solid line), $\alpha=0.1$ (dash line) and $\alpha=0.01$ (dots
line), respectively. From Ref.\cite{Larson:2010gs}, two-dimensional
marginalized
 constraints (68$\%$ and 95$\%$ confidence levels) on inflationary parameters
$r$ and $n_s$, the spectral index of fluctuations, defined at $k_0$
= 0.002 Mpc$^{-1}$. The seven-year WMAP data places stronger limits
on $r$ (shown in red) than five-year data (blue)\cite{WMAP5a}, \cite{WMAP5b}. In
order to write down values that relate $n_s$ and $r$, we used
Eqs.(\ref{ns}) and (\ref{ration}).  Also we have used  the values
$M=-1$, $\epsilon=-1$, $f_1 = 1/2$, $f_2=1$, $\lambda=1/10$ and
$\kappa=1$, respectively.

From Eqs.(\ref{N}), (\ref{ns}) and (\ref{ration}), we observed
numerically that for $\alpha = 1$, the curve $r = r(n_s)$ (see
Fig.\ref{fig2}) for WMAP 7-years enters the 95$\%$ confidence region
where the ratio $r\simeq 0.011$, which corresponds to the number of
e-folds, $N \simeq 32$. For $\alpha=0.1$, $r \simeq 0.103$
corresponds to $N \simeq 227$ and for $\alpha=0.01$, $r \simeq
0.136$ corresponds to $N \simeq 14137$. From 68$\%$ confidence
region for $\alpha = 1$, $r\simeq 0.010$, which corresponds to
$N\simeq 34$. For $\alpha = 0.1$, $r \simeq 0.08$ corresponds to $N
\simeq 240$ and for $\alpha=0.01$, $r \simeq 0.109$ corresponds to
$N \simeq 14279 $.  We noted that the parameter $\alpha$, which lies
in the range $1 > \alpha > 0$, the model is well supported by the
data as could be seen from Fig.\ref{fig2}.

 \section{The vacuum structure of the theory, including the ``kinetic vacuum state" }

 For the discussion of the vacuum structure of the theory, we  start studying $ V_{eff} $ for the case of a constant field  $\phi$, given by,
 \begin{equation}\label{effppluslambda}
 V_{eff}  =
\frac{(f_{1} e^{ \alpha \phi }  +  M )^{2}}{4(\epsilon \kappa ^{2}(f_{1}e^{\alpha \phi}  +  M )^{2} + f_{2}e^{2 \alpha \phi })}+ \Lambda
\end{equation}

 This is necessary, but not enough, since as we will see, the consideration of  constant fields  $\phi$ alone can lead to misleading conclusions, in some cases, the dependence of $ V_{eff} $ on the kinetic term can be crucial to see if and how we can achieve the crossing of an apparent barrier.

For a constant field  $\phi$ the limiting values of $ V_{eff} $ are (now that we added the constant $\Lambda $):

First, for asymptotically
large positive values, ie. as $ \alpha\phi \rightarrow  \infty $,
we have
$V_{eff} \rightarrow
\frac{f_{1}^{2}}{4(\epsilon \kappa ^{2} f_{1}^{2} + f_{2})}+\Lambda $.

Second, for asymptotically large but negative values of the scalar field,
that is as $\alpha \phi \rightarrow - \infty  $ ,  we have:
$ V_{eff} \rightarrow \frac{1}{4\epsilon \kappa ^{2}}+\Lambda = \Delta \lambda $ .

In these two asymptotic regions ($\alpha \phi \rightarrow  \infty  $
and $\alpha \phi \rightarrow - \infty  $) an examination of the scalar
field equation reveals that a constant scalar field configuration is a
solution of the equations, as is of course expected from the flatness of
the effective potential in these regions.

Notice that in all the above discussion it is fundamental that $ M\neq 0$.
If $M = 0$ the potential becomes just a flat one,
$V_{eff} = \frac{f_{1}^{2}}{4(\epsilon \kappa ^{2} f_{1}^{2} + f_{2})}+\Lambda $
everywhere (not only at high values  of $\alpha \phi$).

Finally, there is a minimum at $V_{eff}= \Lambda $ if $M<0$ .
In summary, and if $f_2>0$, $A>0$, we have that there is a hierarchy of vacua ,

\begin{equation}
V_{eff} (\alpha \phi \rightarrow - \infty )=  \Delta \lambda< V_{eff}(min, M < 0)= \Lambda < V_{eff} (\alpha \phi \rightarrow \infty )= C
\end{equation}
where $C = \frac{f_1^2}{4\,f_2(1+\kappa^2\epsilon f_1^2/f_2)}+
\Lambda = \frac{f_1^2}{4f_2}\,A + \Lambda $. notice that we assume
above that $f_1>0$ and $M < 0$, but $f_1<0$ and $M > 0$ would be
indistinguishable from that situation, that is, the important
requirement is $f_1/M < 0$. We could  have a scenario where we start
the non-singular emergent universe at $ \phi \rightarrow \infty $
where $V_{eff} (\alpha \phi \rightarrow \infty )=
\frac{f_{1}^{2}}{4(\epsilon \kappa ^{2} f_{1}^{2} +
f_{2})}+\Lambda$, which then slow rolls, then inflates
\cite{SIchile} and finally gets trapped in the local minimum with
energy density $V_{eff}(min, M < 0)= \Lambda$, that was the picture
favored in \cite{SIchile}, while here we want to argue that the most
attractive and relevant description for the final state of our
Universe is realized after inflation in the flat region $\phi
\rightarrow -\infty$, since in this region the vacuum energy density
is positive and bounded from above, so its a good candidate for our
present state of the Universe. It remains to be  seen however
whether a smooth transition all the way from  $\phi \rightarrow
\infty$ to $\phi \rightarrow -\infty$ is possible.

Before we discuss the transition to the $\phi \rightarrow -\infty$,
it is necessary to discuss another vacuum state, which we may call
the ``kinetic vacuum state" which is in fact degenerate with this
one. The ``kinetic vacuum state" that, with time dependence and say
for no space dependence and $\dot{\phi}^2$ given by

\begin{equation}
\dot{\phi}^2 = - \frac{1}{\epsilon \kappa ^{2}}
\end{equation}

which can be solved for $\dot{\phi}$ in the real domain for $\epsilon < 0$. For this case $R$ (which is not a Riemannian curvature), as given by \ref{e51} diverges, the Riemannian scalar derived from the Einstein frame metric is perfectly regular. In this case then

\begin{equation}
V_{eff}  = \frac{\epsilon R^{2} + U}{(\chi -2 \kappa \epsilon R)^{2} }+ \Lambda \rightarrow  \frac{1}{4\epsilon\kappa^{2}} + \Lambda = \Delta \lambda
\end{equation}
  that is, for this value of $\dot{\phi}^2$, regardless of the value of the scalar field, the value of $V_{eff}$ becomes degenerate with its value for constant and arbitrarily negative $\phi$, which is our candidate vacuum for the present state of the Universe.

  Notice that this value for $\dot{\phi}^2$ is also the one obtained by extremizing the pressure functional in the region of very large scalar field values, so in this limit it is obvious that such configuration satisfy the Euler Lagrange equations, but indeed it is a general feature, the equations of motion for the kinetic vacuum are satisfied, regardless of what value we take for the scalar field.

\section{Evolution of the Universe to its present slowly accelerating state at, crossing ``barriers" and  Dynamical System Analysis.}

In order to discuss the possibility of transition to $\phi
\rightarrow -\infty$ . In our case, since we are interested in a
local minimum between $\phi \rightarrow \infty$ or $\phi \rightarrow
-\infty$, we can take $M$ of either sign.

Taking for definitness $f_1>0$, $f_2>0$, $A>0$,$\epsilon<0$, we see that there will be a point,
given by \ref{effppluslambda}, defined by $\epsilon \kappa ^{2}(f_{1}e^{\alpha \phi}  +  M )^{2} + f_{2}e^{2 \alpha \phi } =0$  where $V_{eff}$ as will spike to $\infty$, go then down to $-\infty$ and then asymptotically its positive asymptotic value at $\phi \rightarrow -\infty$. This has the appearance of a potential barrier. However, this is deceptive, such barrier exists for constant $\phi$, but can be avoided by considering a transition from any $\phi$, but with the appropriate value of $\dot{\phi}^2$ that defines the kinetic vacuum.
A detailed dynamical analysis will be presented now concerning these issues,
The field equations become in the cosmological case:

\begin{eqnarray}
\dot{H} &=& - \frac{k}{2}(\rho + p) - H^2 + \frac{k}{3}\,\rho\,, \label{DS1}\\
\dot{\rho} &=& -3H (\rho + p)\,.\label{DS2}
\end{eqnarray}

By using the definitions of $V(\phi)$ and $U(\phi)$ we can express
$\rho$ and $p$ as follow:

\begin{equation}
\rho = \frac{(1 + \kappa ^{2} \epsilon\,\dot{\phi}^2)\,U}{[U +
\kappa^2\epsilon(V+M)^2]}\,\frac{\dot{\phi}^2}{2} + V_{eff} \,,
\end{equation}

\begin{equation}
p = \frac{(1 + \kappa ^{2} \epsilon\,\dot{\phi}^2)\,U}{[U +
\kappa^2\epsilon(V+M)^2]}\,\frac{\dot{\phi}^2}{2} - V_{eff} \,,
\end{equation}

\begin{equation}
V_{eff} = \frac{(1 + \kappa ^{2}
\epsilon\,\dot{\phi}^2)^2\,(V+M)^2}{4[U +
\kappa^2\epsilon(V+M)^2]^2}\Bigg[ U + \epsilon
\left(\frac{-\kappa(V+M) + \frac{\kappa}{2}\dot{\phi}^2\chi}{(1 +
\kappa ^{2} \epsilon\,\dot{\phi}^2)}\right)^2\Bigg] + \Lambda
\end{equation}

We can note that independent of $\dot{\phi}$ there are a singularity
in the potential (also in $\rho$ and $p$) when $\phi = \phi^*$,
where $U(\phi^*) + \kappa^2\epsilon(V(\phi^*)+M) =0$.

The only way to avoid this situation is consider $\dot{\phi}^2
=-\frac{1}{\kappa^2\, \epsilon}$ before $\phi$ arrive to $\phi^*$.
We can note that, in this case, the effective potential becomes flat
(i.e. independent of $\phi$) and everything is finite.

It is interesting to note that for the case where this model admit
an static and stable universe solution in the region $\phi
\rightarrow \infty$ the kinetic vacuum state solution is an
attractor in the region $\phi >0$, see discussion below.

This situation was already found in the limit
$\phi\rightarrow\infty$ in \cite{SIchile} where the
stability of the static solution was studied.

In particular in the limit $\phi \rightarrow \infty$ the set of
equations (\ref{DS1}, \ref{DS2}) could be written as an autonomous
system of two dimensions respect to $H$ and $y= \dot{\phi}^2$ as
follow, see \cite{SIchile}:

\begin{eqnarray}
\dot{H} &=& \frac{\kappa}{3}\Big[C + B\, y^2 - A(1 +
\kappa^2\,\epsilon\,y)\,y\Big] - H^2, \label{dinamic1}\\
\nonumber \\
\dot{y} &=& - \frac{3A\,(1 + \kappa^2\,\epsilon \,y)\,y}{\frac{A}{2}
+ A\,\kappa^2\,\epsilon \,y + 2B\,y}\,H \,\,, \label{dinamic2}
\end{eqnarray}

As was mentioned in \cite{SIchile} this system has five
critical points where one of these points correspond to the ES
universe discussed previously, but there are also the critical
point

\begin{eqnarray}
H &=& \sqrt{\frac{1}{4\kappa^2\epsilon}+\Lambda} \equiv H_0\\
y &=&-\frac{1}{\kappa^2\, \epsilon} \equiv y_0\,.
\end{eqnarray}

\begin{figure}
\centering
\includegraphics[width=7cm]{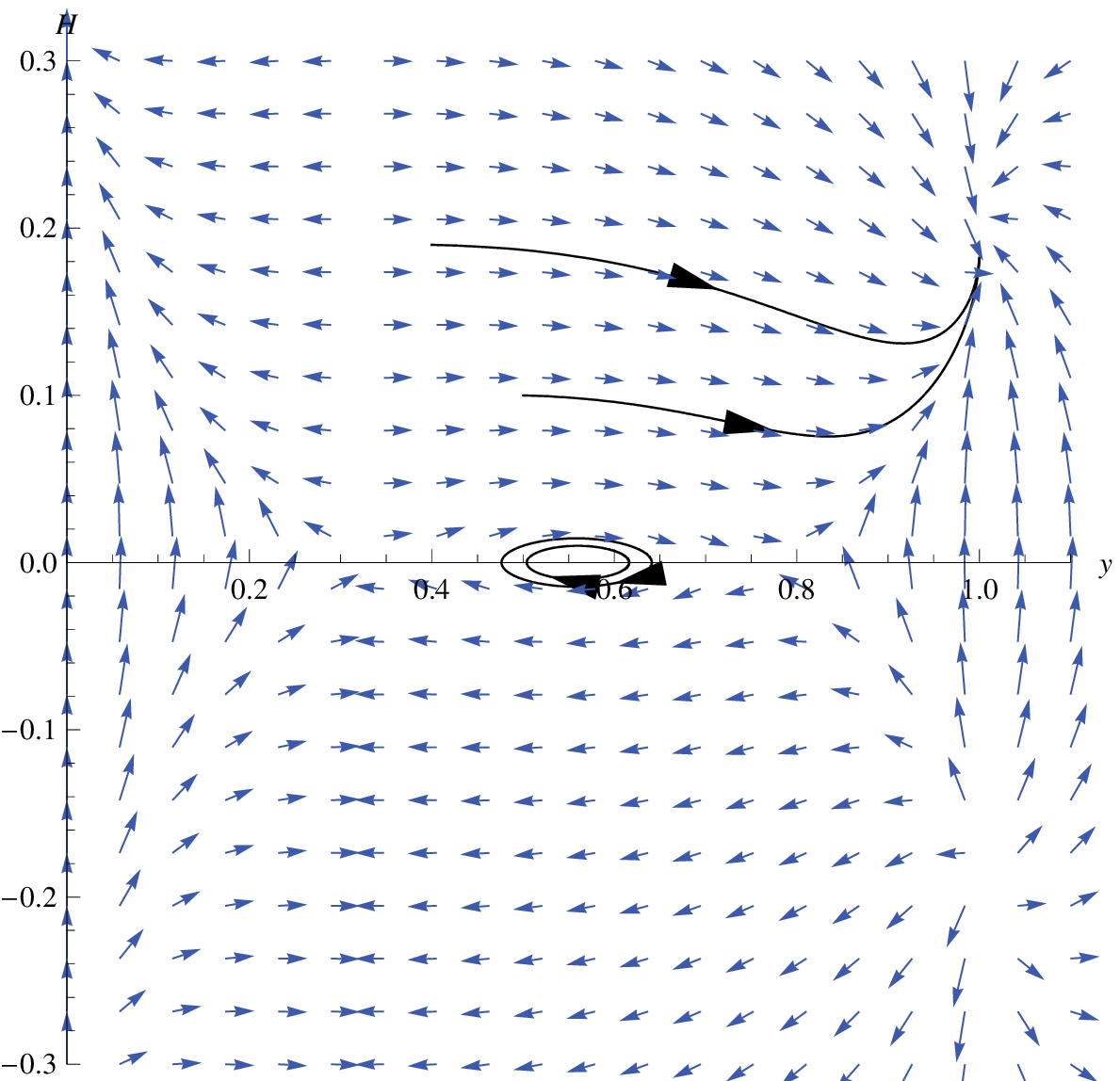}
\includegraphics[width=7cm]{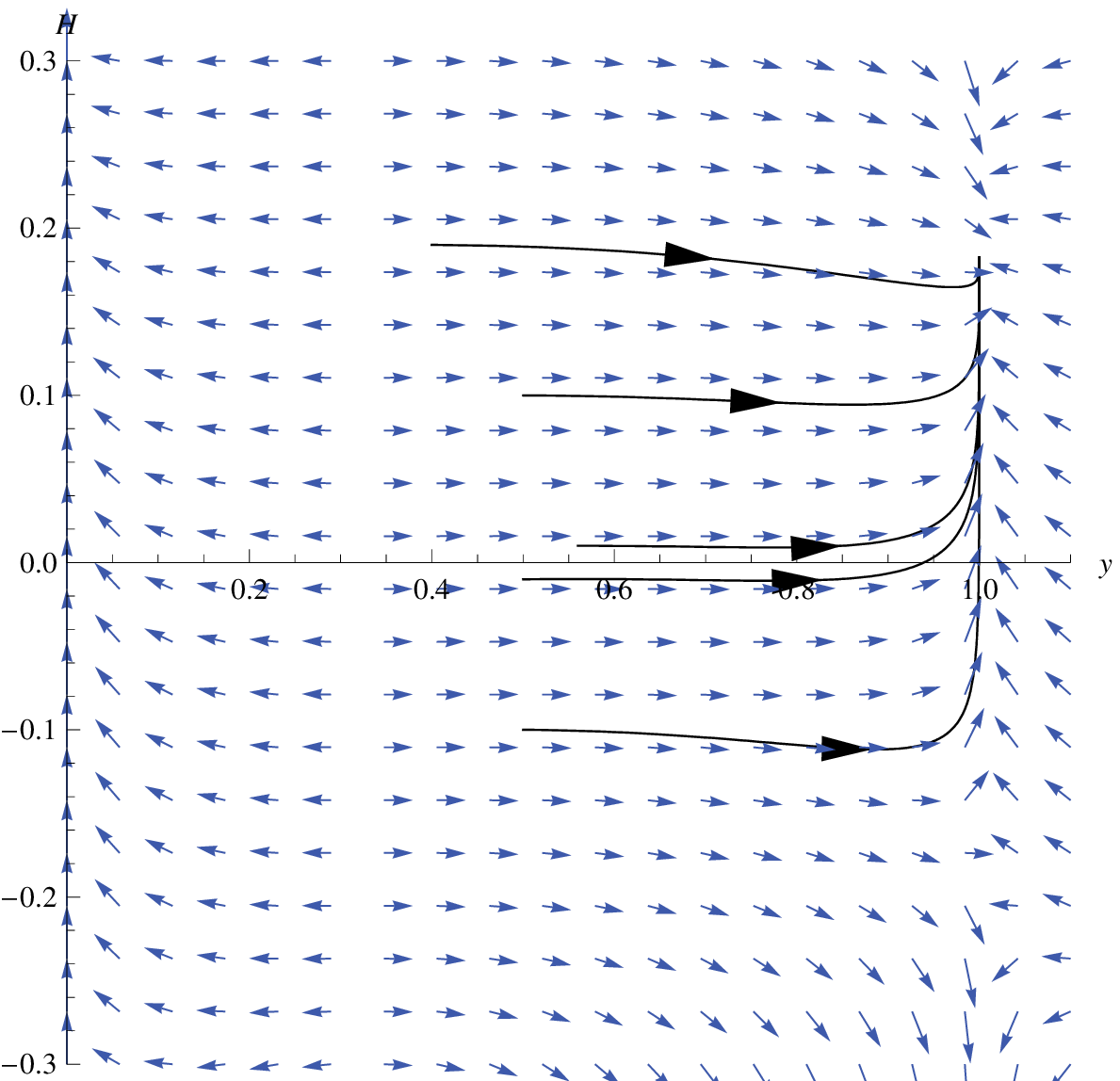}
\includegraphics[width=7cm]{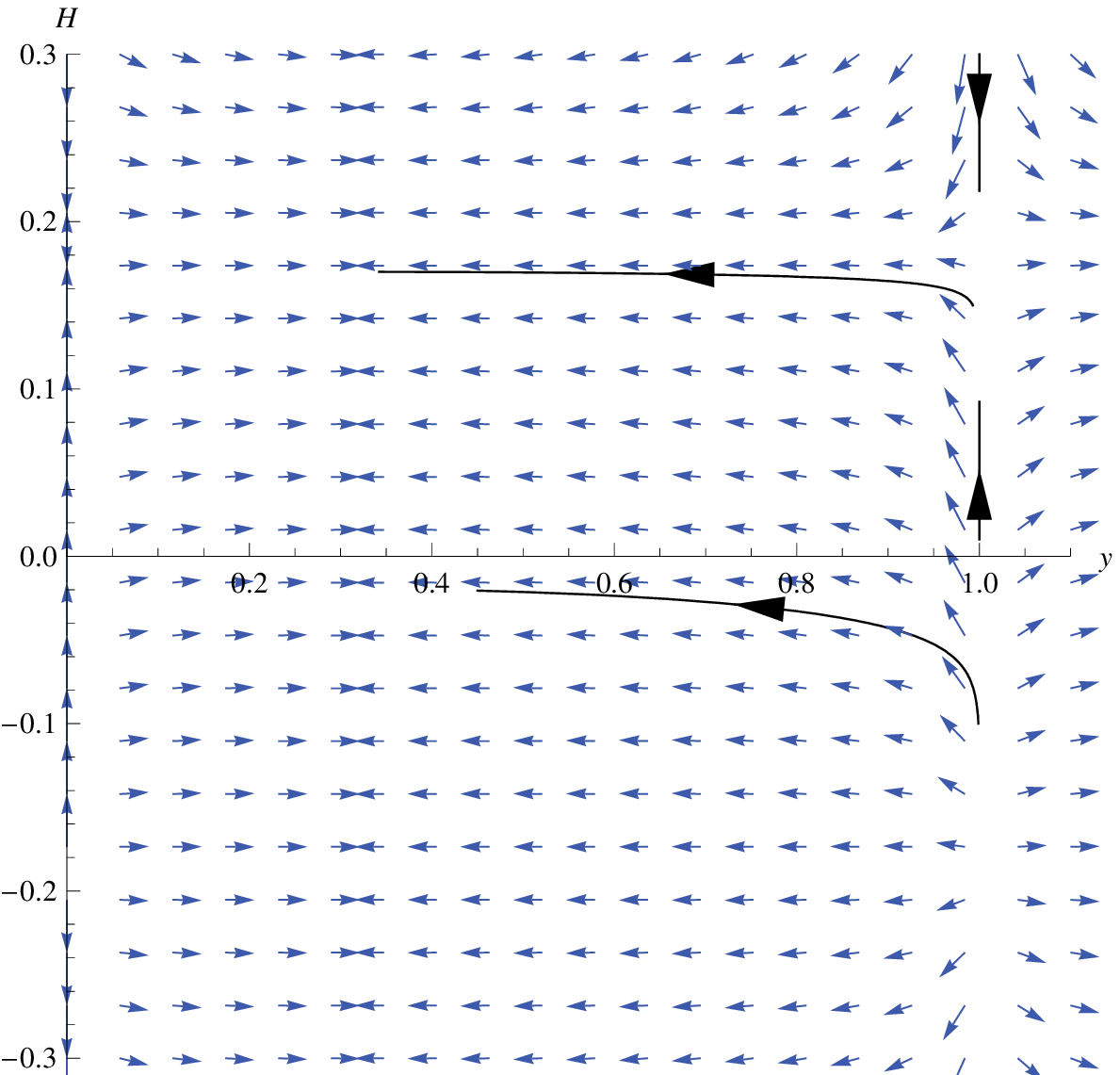}
\caption{Plots showing part of the \textit{Direction Field} of the
system for different values of $\phi$ and some numerical
solutions.\label{FDS1}}
\end{figure}

This critical point is, precisely, the kinetic vacuum state which
avoid the singularity problem of $V_{eff}$ discussed above.
After we linearize the equations (\ref{dinamic1}, \ref{dinamic2}) near
this critical point we obtain that the eigenvalues of the linearized
equations are negative $\lambda_1= -2H_0$ and $\lambda_2=-3H_0$,
then, this critical point is an attractor. The Fig. (\ref{FDS1}) top
left panel  show part of the \textit{Direction Field} of the system
and four numerical solutions where we can note that the kinetic
vacuum state is an attractor solution.

It is interesting to note that this solution is in fact an attractor
not only in the limit $\phi \rightarrow \infty$, but also in others
regions.

In order to study this point in more details let us write the set of
equations (\ref{DS1}, \ref{DS2}) as an autonomous system of three
dimensions as follow:

\begin{eqnarray}
\dot{H} &=& - H^2 + \frac{\Lambda}{3} +
\frac{\kappa}{12}\frac{(M+V)^2 - U\,y\,(4+3\kappa^2\,\epsilon\,y)}{U
+ \kappa^2\,\epsilon\,(M+V)^2}\;, \label{SDC1}\\
\nonumber \\
 \dot{y} &=& \frac{1 + \kappa^2\,\epsilon\,y}{1 +
3\kappa^2\epsilon\,y}\, \Bigg(-6H\,y + \frac{\alpha\,M(M+V)(-1 +
3\kappa^2\epsilon\,y)\sqrt{y}}{U + \kappa^2\,\epsilon(M+V)^2}
\Bigg)\;,\label{SDC2}\\
\nonumber \\
\dot{\phi} &=& -\sqrt{y}\label{SDC3}
\end{eqnarray}

\begin{figure}
\centering
\includegraphics[width=7cm]{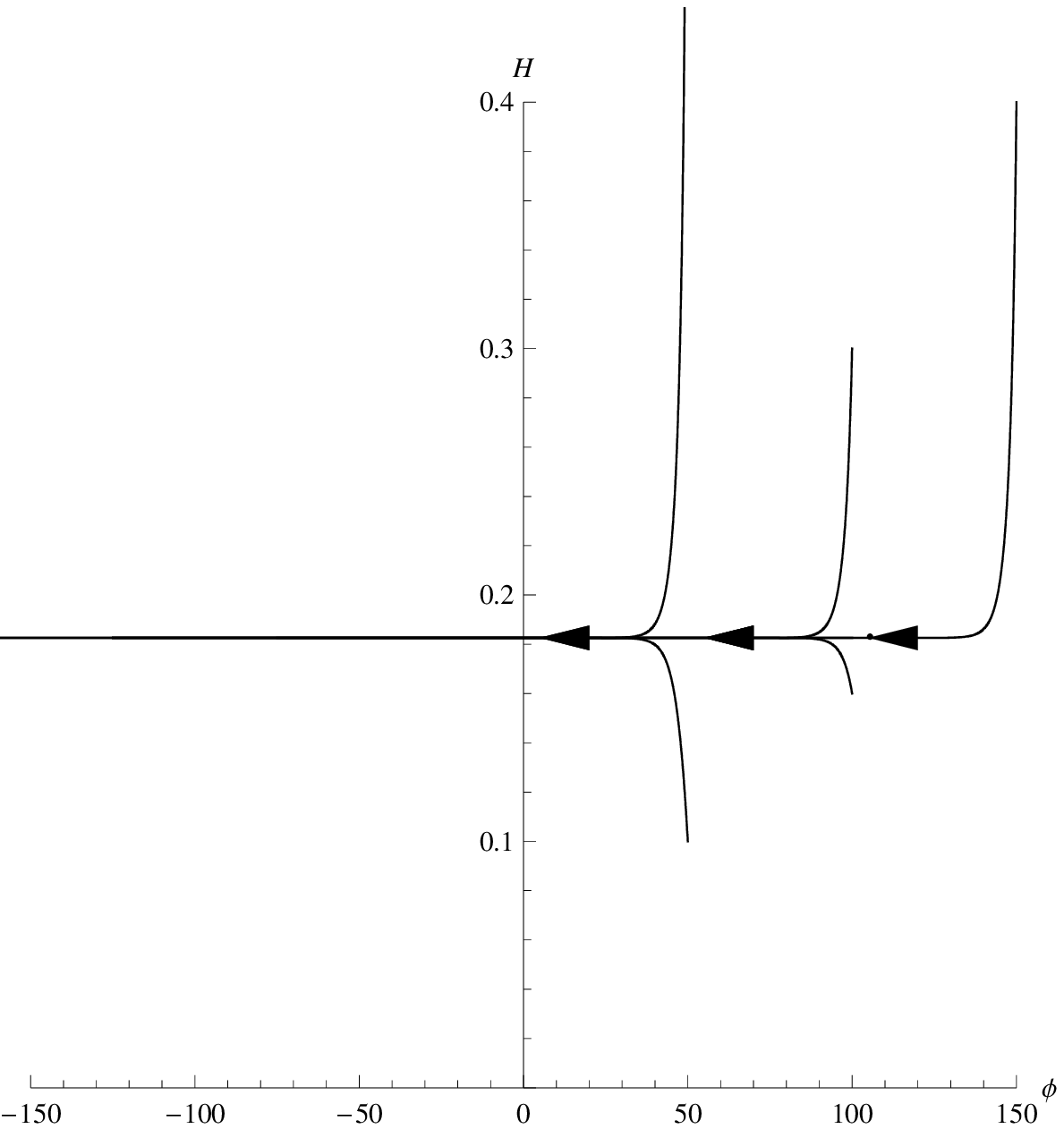}
\includegraphics[width=7cm]{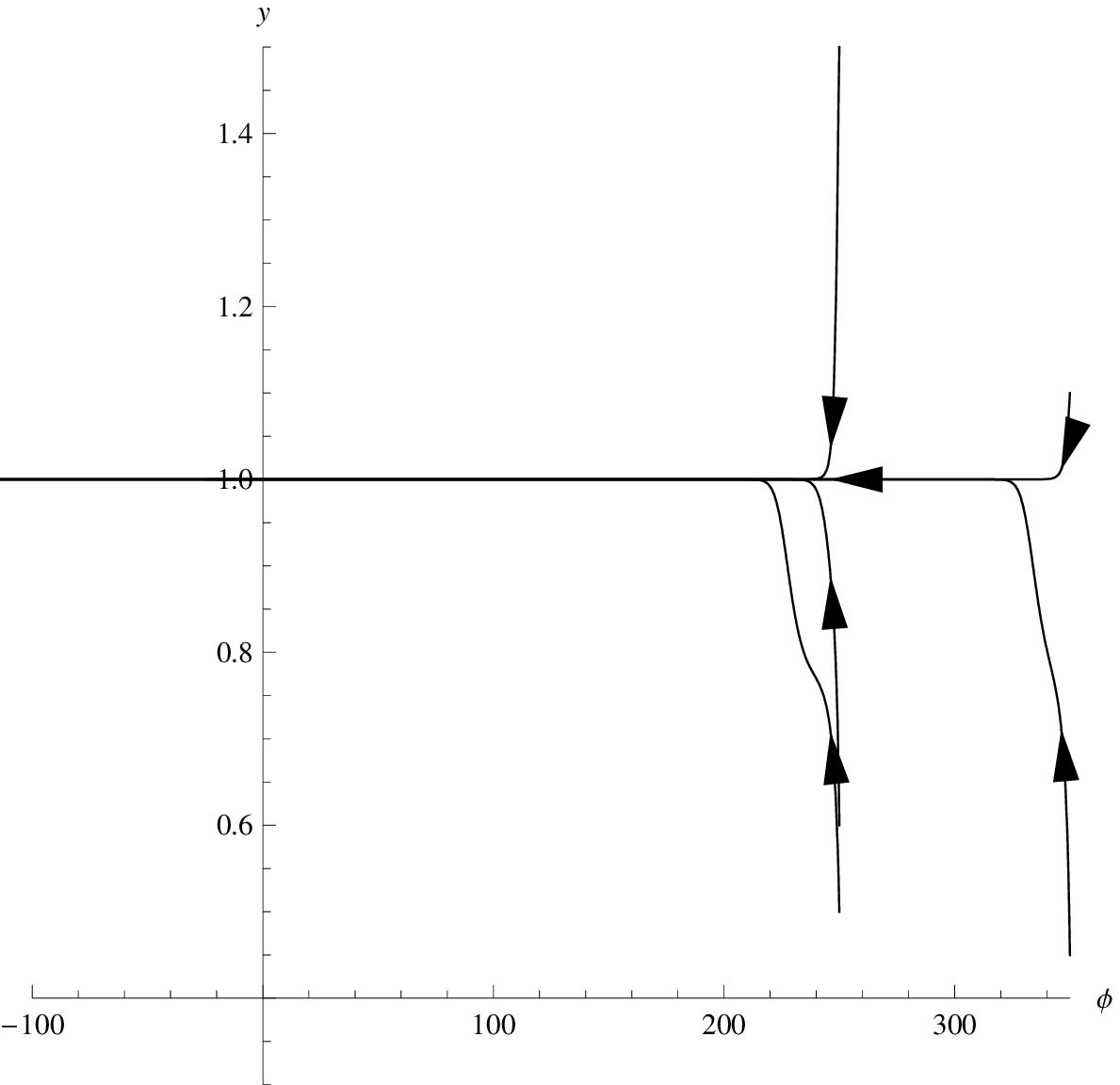}
\caption{Plots showing some numerical solution of the
Eqs.(\ref{SDC1}, \ref{SDC2}, \ref{SDC3}). \label{FNS1}}
\end{figure}

Where we have defined $y = \dot{\phi}^2$. We are consider
$\dot{\phi} <0$ because we are interested in the cases where the
field moves from $-\infty$ to positive values, following the
Emergent Universe scheme. We can can note that, in general, the
solutions $H = H_0$, $y=y_0$ is stable. This solution correspond to
a flat effective potential and $\phi$ rolling with constant
$\dot{\phi}$. In this case, we can past over the point
$\phi=\phi^*$, see numerical solutions Fig.(\ref{FNS1}). Also, we
can observed that solutions near the kinetic vacuum solution can
pass over this point , because this solution is an attractor.
The general behaviour could be see in the Fig. (\ref{FDS1}) where it
is plot the \textit{Direction Field} for the effective two
dimensional autonomous system in variables $H$ and $y$ which we
obtain when evaluate the system of Eqs.(\ref{SDC1},
\ref{SDC2},\ref{SDC3}) at different values of $\phi$. The first plot
correspond to the limit $\phi \rightarrow\infty$ the second is for
$\phi \gtrsim \phi^*$, and  the third correspond to $\phi \ll
\phi^*$.

In order to study in a more systematic way the nature of the kinetic
vacuum state we linearize the Eqs.(\ref{SDC1},
\ref{SDC2},\ref{SDC3}) near the critical point $H_0, y_0$ leaving
$\phi$ arbitrary. We obtain the following two dimensional effective
autonomous system with variables $\delta H$ and $\delta y$:

\begin{eqnarray}
\delta \dot{H}= - 2H_0\,\delta H + \frac{\chi\,\delta
y}{12\epsilon\Big[\kappa (V+M)+\frac{\chi}{2\kappa\,\epsilon}\Big]}
\,,\label{Pert1}\\
\nonumber \\
\delta \dot{y}= \Bigg(-3H_0 +
\frac{2\alpha\,M\,\epsilon\,\kappa^2(M+V)\sqrt{y_0}}{U +
\kappa\,\epsilon(M+V)^2}\Bigg)\delta y\,.\label{Pert2}
\end{eqnarray}

The eigenvalues of equations (\ref{Pert1}), (\ref{Pert2}) are:

\begin{eqnarray}
\lambda_1 &=& -2H_0 \,,\\
\lambda_2 &=& -3H_0 +
\frac{2\alpha\,M\,\epsilon\,\kappa^2(M+V)\sqrt{y_0}}{U +
\kappa\,\epsilon(M+V)^2} \,.
\end{eqnarray}

The equilibrium point is stable (attractor) if the eigenvalues are
negative. Then, depending on the values of the parameters of the
models, this is the case for a large set of values of $\phi$, not
only for the case $\phi \rightarrow +\infty$ discussed previously in
\cite{SIchile}. In particular for the numerical values used
in \cite{SIchile}, which are consistent with the stability
of the ES universe, the critical point is stable for $\phi >
\phi^*$, see Fig. (\ref{FDS1}).

It is interesting to note that when $\phi \rightarrow -\infty$, then

\begin{equation}
\lambda_2 \rightarrow -3H_0 + 2\alpha\,\sqrt{y_0} \,.
\end{equation}

For example, if we consider the numerical values used in
\cite{SIchile} this is a positive number. This means that at
some value of $\phi < \phi^*$, when the scalar field moves to
$-\infty$ the stable (attractor) equilibrium point becomes unstable
(Focus), see Fig. (\ref{FDS1}).

In order to study numerical solutions we chose the following values
for the free parameters of the model, in units where $\kappa = 1$;
$f_1= 1$, $f_2=1/2$, $\epsilon =-1$, $\alpha=1$, $\Lambda = 0.35$
and $M=-1$. These values satisfy the requirements of stability of
the ES solution in the limit $\phi >> 0$, see
Ref.\cite{SIchile}. Under this assumptions we obtain that:

\begin{eqnarray}
\phi^* &=& -0.40 \\
H_0 &=& 0.18 \\
y_0 &=& 1
\end{eqnarray}

Numerical solution to the Eqs.(\ref{SDC1}, \ref{SDC2}, \ref{SDC3})
are show in Fig.(\ref{FNS1}), where we can note that the point
$\phi=\phi^*$ is passed on during the evolution of the scalar field.
This situation is achieved by the kinetic vacuum solution, but also
by others solutions which decays to the kinetic vacuum solution
before arrive to the point $\phi=\phi^*$, see Fig. (\ref{FNS1}).

The figure (\ref{FNS1}) left panel shown a projection of the axis
$H$ and $\phi$ and the evolution of six numerical solutions. The
right panel shown a projection of the axis $y$ and $\phi$ and the
evolution of six numerical solutions.


\section{Discussion, the creation of the universe as a ``threshold event" for zero present vacuum energy density, when does the bound restrict us to a small vacuum energy density for the late universe? }

We have considered a non-singular origin for the Universe starting
from an Einstein static Universe, the so called ``emergent universe"
scenario, in the framework of a theory which uses two volume
elements  $\sqrt{-{g}}d^{4}x$ and $\Phi d^{4}x$, where $\Phi $ is a
metric independent density, used as an additional measure of
integration. Also curvature, curvature square terms and for scale
invariance a dilaton field $\phi$  are considered in the action. The
first order formalism was applied.  The integration of the equations
of motion associated with the new measure gives rise to the
spontaneous symmetry breaking (S.S.B) of scale invariance (S.I.).
After S.S.B. of S.I., using the the Einstein frame metric, it is
found that a non trivial potential for the dilaton is generated. One
could question the use of the Einstein frame metric
$\overline{g}_{\mu\nu} $ in contrast to the original metric
$g_{\mu\nu} $ . In this respect, it is interesting to see the role
of both the original metric and that of the Einstein frame metric in
a canonical approach to the first order formalism. Here we see that
the original metric does not have a canonically conjugated momentum
(this turns out to be zero), in contrast, the canonically conjugated
momentum to the connection turns out to be a function exclusively of
$\overline{g}_{\mu\nu} $, this Einstein metric is therefore a
genuine dynamical canonical variable, as opposed to the original
metric.

There is also a lagrangian formulation of the theory which uses
$\overline{g}_{\mu\nu} $, what we can call the action in the
Einstein frame. In this frame we can quantize the theory for example
and consider contributions without reference to the original frame,
thus possibly considering breaking the TMT structure of the theory,
but such breaking will be done ``softly" through the introduction of
a cosmological term only. Surprisingly, the remaining structure of
the theory, reminiscent from the original TMT structure will be
enough to control the strength of this additional cosmological term
once we demand that the universe originated from a non-singular and
stable emergent state.

In the Einstein frame we argue that the cosmological term  parametrizes the zero point fluctuations.

The resulting effective potential for the dilaton contains two flat
regions, for $\phi \rightarrow \infty$ relevant for the non-singular
origin of the Universe, followed by an inflationary phase and then
transition to $\phi \rightarrow -\infty$, which in this paper we
take as describing our present Universe. An intermediate local
minimum is obtained if $f_1/M<0$, the region as $\phi \rightarrow
\infty$ has a higher energy density than this local minimum and of
course of the region $\phi \rightarrow -\infty$, if  $A>0$ and $f_2
>0$. $A>0$ is also required for satisfactory slow roll in the
inflationary region $\phi \rightarrow \infty$ (after the emergent
phase). The dynamics of the scalar field becomes non linear and
these non linearities are instrumental in the stability of some of
the emergent universe solutions, which exists for a parameter range
of values of the vacuum energy in $\phi \rightarrow -\infty$, which
must be positive but not very big, avoiding the extreme fine tuning
required to keep the vacuum energy density of the present universe
small. A sort of solution of the Cosmological Constant Problem,
where an a priori arbitrary cosmological term is restricted by the
consideration of the non-singular and stable emergent origin for the
universe.

Notice then that the creation of the universe can be considered as a
``threshold event" for zero present vacuum energy  density, that is
a threshold event for  $\Delta \lambda = 0$ and we can learn what we
can expect in this case by comparing with well known threshold
events. For example in particle physics, the process $e^{+}+e^{-}
\rightarrow \mu^{+}+\mu^{-}$, has a cross section of the form
(ignoring the mass of the electron and considering the center of
mass frame, $E$ being the center of mass energy of each of the
colliding $e^{+}$ or $e^{-}$),

\begin{equation}
 \sigma_{e^{+}+e^{-} \rightarrow \mu^{+}+\mu^{-}} = \frac{\pi \alpha^{2}}{6 E^{2} }\left[ 2 + \frac{m_{\mu}^2}{E^2}  \right]\sqrt{\frac{E^{2}-m_{\mu}^2}{ E^{2}}}
\end{equation}
for $E>m_{\mu}$ and exactly zero for $E < m_{\mu}$ . We see that exactly at threshold this cross section is zero, but at this exact point it has a cusp, the derivative is infinite and the function jumps as we slightly increase $E$.
By analogy, assuming that the vacuum energy can be tuned somehow (like the center of mass energy $E$ of each of the colliding particles in the case of the annihilation process above), we can expect zero probability for exactly zero vacuum energy density $\Delta\lambda =0$, but that  soon after we build up any positive $\Delta\lambda$ we will then able to create the universe, naturally then, there will be a creation process resulting in a universe with a small but positive $\Delta\lambda$ which represents the total energy density for the region describing the present universe, $\phi \rightarrow -\infty$ or by the kinetic vacuum (which is degenerate with that state).

One may ask the question: how is it possible to discuss the
``creation of the universe" in the context of the ``emergent
universe"?. After all, the Emergent Universe basic philosophy is
that the  universe had a past of infinite duration. However, that
most simple notion of an emergent universe with a past of infinite
duration has been recently challenged by Mithani and Vilenkin
\cite{Mithani1}, \cite{Mithani2} at least in the context of a
special  model. They have shown that an emergent universe, although
completly stable classically, could be unstable under a tunnelling
process to collapse. On the other hand, an emergent universe can
indeed be created from nothing by a tunnelling process as well.

An emerging universe could last for a long time provided it is
classically stable, that is where the constraints on the
cosmological constant for the late universe discussed here come in.
If it is not stable, the emergent universe will not provide us with
an appropriate ``intermediate state" connecting the creation of the
universe with the present universe. The existence of this stable
intermediate state provides in our picture the reason for the
universe to prefer a very small vacuum energy density at late times,
since universes that are created, but do not make use of the
intermediate classically stable emergent universe will almost
immediately recollapse, so they will not be ``selected". Finally, it
could be that we arrive to the emergent solution not by quantum
creation from nothing, by the evolution from something else, for
example by the production of a bubble in a pre-existing state
\cite{Labrana}, from here we go on to inflation.

Notice that the bound gives a small vacuum energy density, without reference to the threshold mechanism mentioned before. For this notice that upper the bound on the present vacuum energy density of the universe contains a $1/\epsilon$ suppression. If we think of the $R^2$ term as generated through radiative corrections, $\epsilon$ is indeed formally infinite, in dimensional regularization
goes as $\epsilon = K/(D-4)$, \cite{hooft1}-\cite{hooft3} so it can have either sign (depending on how we approach $D-4=0$). In any case, a very large $\epsilon$ means a very strict bound on the present vacuum energy density of the universe.

\section{Acknowledgements}

We would like to thank Sergio del Campo and Ramon Herrera, our
coauthors in ref. \cite{SIchile}, which is central for our review,
for our crucial collaboration with them and for discussions on all
the subjects in this review, in particular we have benefited from
our discussions concerning the aspects related to inflation, slow
roll, etc. in the context of the model studied here. We also want to
thank Zvi Bern, Alexander Kaganovich and Alexander Vilenkin for very important
additional discussions. UCLA, Tufts University and the University of
Barcelona are thanked for their wonderful hospitality. P. L. has
been partially supported by FONDECYT grant N$^{0}$ 11090410, Mecesup
UBB0704 and Universidad del B\'{i}o-B\'{i}o through grant DIUBB
121407 GI/VC.

\break

\end{document}